\documentclass[11pt,a4paper]{article}
\pdfoutput=1

\usepackage[]{Packages}
\RequirePackage{placeins}
\RequirePackage{ragged2e}
\newcommand{\W}{\mathcal{W}}
\usepackage{Definitions}
\usepackage[mathscr]{euscript}
\allowdisplaybreaks[1]

\newcommand{\OfficialTitle}{Locality and anomalies in warped conformal field theory}

\title{\vspace{2cm}
  {\huge   \textbf{\OfficialTitle}}
}

\author{
  \begin{minipage}{.8\linewidth}
    \vspace{1cm}
    \begin{center}
      {\small \textbf{Kristan Jensen}}
    \end{center}
    \vspace{1cm}
    \begin{minipage}{\linewidth}
      {\itshape \footnotesize \begin{center}
        San Francisco State University \\ \vspace{.1cm}
        San Francisco, CA 94132 \end{center}
      }
    \end{minipage}
    \vspace{2cm}
  \end{minipage}
}

\date{\today}

\begin{document}

\setstretch{1.1}

\numberwithin{equation}{section}

\begin{titlepage}

  \maketitle

  \thispagestyle{empty}


  \abstract{\RaggedLeft We study various aspects of warped conformal field theories (WCFTs) in two dimensions. We find new Lagrangian WCFTs, and show that all known Lagrangian WCFTs are local only after an infinite tuning. We deduce the anomalies of WCFTs and relate them to central charges, as well as analyze WCFT hydrostatics, thereby re-deriving the warped analogue of the Cardy formula. Finally, we point out that holographic WCFTs are semi-local, that matter fields in spacelike WAdS$_3$ black hole geometries exhibit a Gregory-Laflamme-like instability, and that WAdS$_3$ throats do not admit a decoupling limit.}
\end{titlepage}

\clearpage

\numberwithin{equation}{section}
\newcommand{\beq}{\begin{equation}}
\newcommand{\eeq}{\end{equation}}
\newcommand{\M}{\mathcal{M}}

\def\KJ#1{\textbf{[KJ:#1]}}

\tableofcontents

\section{Introduction} 

Extremal charged black branes possess a near-horizon AdS$_{d+1}\times \mathcal{M}$ throat. Under favorable circumstances, physics in the throat decouples at low energies, leading to two consequences. First, from a more practical point of view, the decoupling simplifies the computation of low-energy $S$-matrix elements, owing to the high symmetry of AdS$_{d+1}$. Second, it is generally believed that gravitational physics in the AdS throat is dual to a $d$-dimensional conformal field theory (CFT).

Extremal rotating black holes have a more complicated near-horizon, known as the near-horizon extremal Kerr (NHEK) geometry. In four dimensions NHEK is topologically AdS$_2\times \mathbb{S}^2$, and its high degree of symmetry also leads to a simplification of low-energy $S$-matrix elements. This near-horizon is the main actor in the ``Kerr/CFT'' correspondence of~\cite{Guica:2008mu}, the conjecture that quantum gravity on the NHEK geometry has a CFT dual.

At fixed polar angle in the four-dimensional NHEK geometry, one finds an interesting three-dimensional geometry known as warped AdS$_3$ or WAdS$_3$. (It is analogous to a warped $\mathbb{S}^3$, in which the Hopf fiber is either stretched or squashed.) WAdS$_3$ spacetimes have $PSL(2;\mathbb{R})\times U(1)$ isometry, and with suitable boundary conditions, the asymptotic symmetry group of WAdS$_3$ is enhanced from $PSL(2;\mathbb{R})\times U(1)$ to a Virasoro-Kac-Moody symmetry. In the wake of the Kerr/CFT correspondence, it has also been conjectured that quantum gravity on WAdS$_3$ spacetimes has a field theory dual, dubbed a ``warped conformal field theory'' or WCFT.

The asymptotic symmetries of WAdS$_3$ do not include a boundary Lorentz boost, and so if a dual WCFT exists, it is not Lorentz-invariant. Thus a putative WCFT/WAdS$_3$ correspondence would be an example of non-relativistic holography.

What is a WCFT? One definition is that it is a two-dimensional theory invariant under ``warped conformal symmetries.'' Parameterizing the plane with coordinates $x^{\pm}$, which we refer to as left and right-moving, the warped conformal symmetries are given by
\beq
\label{E:warpedSymmetries}
x^- = x^-(X^-)\,, \qquad x^+ = X^+ + f(X^-)\,,
\eeq
and are generated by a right-moving stress tensor $T(x^-)$ and current $P(x^-)$ respectively, leading to a Virasoro-Kac-Moody algebra, rather than the Virasoro$^2$ symmetry of an ordinary $2d$ CFT. Observe that the transformations of the left-moving direction $x^+$ are generated by the right-moving current $P(x^-)$. The warped symmetries~\eqref{E:warpedSymmetries} exactly match the asymptotic symmetry group of WAdS$_3$ subject to the boundary conditions of~\cite{Compere:2008cv}. 

There has been significant study of both WAdS$_3$ spacetimes and WCFTs. See e.g.~\cite{Anninos:2008fx,Compere:2008cv,Anninos:2008qb,Guica:2010sw,Hofman:2011zj,ElShowk:2011cm,Song:2011sr,Guica:2011ia,Azeyanagi:2012zd,Detournay:2012pc,Compere:2013bya,Anninos:2013nja,Hofman:2014loa,Hartong:2015xda,Castro:2015uaa,Castro:2015csg,Castro:2017mfj,Guica:2017lia}. A few results are worth mentioning here. First, using the warped conformal symmetries, Detournay, Hartman, and Hofman~\cite{Detournay:2012pc} have derived a warped analogue of the Cardy formula. Their result expresses the asymptotic density of states of a general WCFT in terms of its central charges and the vacuum charges of $L_0$ and $J_0$ on a circle. The logarithm of this density of states is the high-temperature thermal entropy, expressed as a function of the charges, and this entropy matches that of WAdS$_3$ black holes. Further, there is a universal expression~\cite{Castro:2015csg} for the single-interval entanglement entropy of a WCFT.

Hofman and Rollier~\cite{Hofman:2014loa} constructed the two known examples of Lagrangian WCFT to date, namely the warped analogue of a Weyl fermion and a warped $bc$ theory. They also suggested that, just as $2d$ CFTs naturally couple to Riemannian geometry, WCFTs couple to a structure they dubbed ``warped geometry.'' This structure is a variant of Newton-Cartan geometry, which has received some attention of late in the context of Son's work on the fractional quantum Hall effect~\cite{Geracie:2014nka} as well as in non-relativistic holography~\cite{Christensen:2013rfa}.

There is an expectation that WCFTs are generically non-local. Before the Hofman/Rollier theories, the known examples of WCFTs were the IR limits of dipole-deformed $2d$ CFTs as in~\cite{ElShowk:2011cm,Song:2011sr,Azeyanagi:2012zd}. When the parent $2d$ CFT has a gravity dual, then after turning on the dipole deformation, the bulk geometry has a WAdS$_3$ factor. The dipole deformation breaks locality, and it is not a priori clear whether locality is restored in the IR. 

Thinking of $x^-$ as ``time,'' it is clear from~\eqref{E:warpedSymmetries} that WCFTs are non-relativistic theories with a dynamical critical exponent $z=\infty$ (meaning that under dilatations time rescales but space does not). Locality is precarious in any theory with $z=\infty$, as spatial derivatives do not carry any dimension, and from this alone one might expect that a generic WCFT is non-local. Due to their high degree of symmetry, WCFTs would then offer perhaps the most controlled examples of non-local field theories.

In this paper we investigate various aspects of WCFT, focusing on the role of locality and anomalies. We establish a number of loosely related results. After reviewing preliminary material in Section~\ref{S:review}, we write down two new Lagrangian WCFTs in Section~\ref{S:freeWCFT}. They are free scalar analogues of the Hofman/Rollier theories. Basically because WCFTs have a dynamical exponent $z=\infty$, all of theories possess exactly marginal, non-local deformations quadratic in the fields. In fact there are an infinite number of such operators, and all must be tuned away to obtain a local WCFT.

Next in Section~\ref{S:anomalies} we use Wess-Zumino consistency to determine the anomalies of local WCFTs. We find three anomalies. One is the analogue of the usual trace anomaly of $2d$ CFT, and is governed by the Virasoro central charge. The second is a pure boost anomaly, which does not have a $2d$ CFT analogue, and it is governed by the Kac-Moody central charge. The third is a gravitational anomaly, which is unrelated to the Virasoro and Kac-Moody central charges.

Much like in $2d$ CFT, there is a warped analogue of the conformal transformation from the infinite plane to the thermal cylinder. In Section~\ref{S:thermodynamics} we use this result (combined with the hydrostatic partition function technology of~\cite{Jensen:2012jh,Banerjee:2012iz}) to obtain the high-temperature thermodynamics of any local WCFT. Our result depends on the anomaly coefficients and on the one-point functions of $T$ and $P$ on the plane, and we recover the warped Cardy formula of Detournay, Hartman, and Hofman.

Finally we turn our attention to holographic WCFTs, or at least, the putative duals to theories of gravity on spacelike WAdS$_3$ spacetimes. See Section~\ref{S:holography}. We do not consider the ``lower-spin gravity'' of~\cite{Hofman:2014loa}. Spacelike WAdS$_3$ spacetimes have a spatial circle fibered over AdS$_2$, and the fibration is characterized by a constant $\alpha$. When $\alpha=1$ one has AdS$_3$, while for $\alpha>1$ the circle is ``stretched.'' We find that spacelike WAdS$_3$ behaves rather like the AdS$_2\times \mathcal{M}$ near-horizon geometry of charged black holes and black branes, and in particular, the putative dual is semi-local in the sense of~\cite{Iqbal:2011in}.\footnote{For WCFTs realized as the IR limits of dipole-deformed $2d$ CFTs with an AdS dual, this shows that the non-locality persists in the IR.} A scalar field on spacelike WAdS$_3$ cannot be dual to a local operator. Using the standard holographic dictionary, Fourier modes around the spatial circle of WAdS$_3$ with momentum $k$ are dual to boundary operators whose conformal dimension depends on $k$. (Schr\"odinger holography has similar features~\cite{Guica:2010sw}.) We find a number of other results. Spacelike WAdS$_3$ black holes exist when $\alpha>1$, however, we find that in this regime minimally coupled matter fields exhibit a linearized instability even for pure WAdS$_3$. Modes which carry momentum $k$ along the circle are unstable for $|k|>k_c$, analogous to the Gregory-Laflamme instability. 

Ignoring this instability, there is another issue with matter fields, namely even their $k=0$ modes lead to strong backreaction that destroys the WAdS$_3$ asymptotics, just like for matter fields in an AdS$_2$ geometry with a finite volume transverse space. So WAdS$_3$ throats with a finite volume spatial circle never truly decouple. With the non-locality in mind, perhaps the best way to think of the dual to gravity on WAdS$_3$ is as a large $N$ quantum mechanics with an approximate low-energy conformal symmetry and a $U(1)$ global symmetry, like the complex Sachdev-Ye-Kitaev models of recent interest.

With the rest of the paper in mind, we also argue that the conformal boundary of spacelike WAdS$_3$ is equipped with the warped geometry of~\cite{Hofman:2014loa} subject to a constraint that we discuss. Moreover the anomalies we uncover in Section~\ref{S:anomalies} are not matched by any gravitational theory on WAdS$_3$. This is puzzling. In a local WCFT, the Virasoro and Kac-Moody central charges are determined by the anomalies, but the authors of~\cite{Compere:2008cv} have obtained Virasoro-Kac-Moody symmetry (with nonzero central charges) from gravity. So here it seems that there is a Virasoro-Kac-Moody symmetry without corresponding anomalies. We suspect that the resolution to this puzzle is that while anomalies are tied to central charges in a local theory, perhaps there is no such relation in the absence of locality, in which case there is no inconsistency.

\section{Preliminaries} 
\label{S:review}

\subsection{WCFTs in flat space}
\label{S:flatWCFT}

One starting point for the definition of a local WCFT is the following~\cite{Hofman:2011zj}. Consider a unitary, local quantum field theory in two dimensions; labeling the coordinates of the plane as $x^{\pm}$,\footnote{Despite the suggestive name of the coordinates, we do not presume a Lorentzian structure here or elsewhere. However we will refer to $x^-$ as a ``right-moving'' direction in analogy with the Lorentzian plane.} we assume that the theory is invariant under translations and ``right-moving'' dilatations
\beq
\label{E:basicSymmetries}
x^{\pm} \to x^{\pm} + c^{\pm}\,, \qquad x^-\to \lambda x^-\,.
\eeq
As shown in~\cite{Hofman:2011zj}, unitarity, locality, and a bounded below spectrum of the dilatation operator together imply that translations and dilatations are enhanced to an infinite-dimensional group of symmetries. The enhancement depends on the theory. One possibility is that the symmetry algebra enhances to two copies of the Virasoro algebra, in which case one has a conventional CFT. Another is that right dilatations and translations enhance to a right-moving Virasoro algebra, while left- translations are enhanced to a right-moving abelian Kac-Moody algebra. This second case is a WCFT. 

The global subgroup of the warped conformal symmetries is $PSL(2;\mathbb{R})\times \mathbb{R}$ when $x^+$ is non-compact, and $PSL(2;\mathbb{R})\times U(1)$ when $x^+$ is compact.

The Virasoro and Kac-Moody symmetry currents generate the transformations~\cite{Hofman:2011zj}
\beq
\label{E:WCFTsymmetries}
x^- = x^-(X^-)\,, \qquad x^+ = X^+ + f(X^-)\,.
\eeq
There is a stress tensor $T(x)$ which generates the reparameterizations of $x^-$, and a momentum density $P(x)$ which generates the additive reparameterizations of $x^+$. Compactifying $x^-$, the Virasoro algebra is generated by the modes of $T(x)$ and the Kac-Moody by those of $P(x)$. The Kac-Moody is a $U(1)$ algebra when $x^+$ is compact -- charged states carry momentum in the $+$ direction -- and an $\mathbb{R}$ algebra when $x^+$ is non-compact. The two-point functions of $T(x)$ and $P(x)$ on the infinite plane are
\beq
\langle T(x)T(0)\rangle = \frac{c}{2(x^-)^4}\,, \qquad \langle P(x) P(0)\rangle = \frac{k}{(x^-)^2}\,.
\eeq
$c$ and $k$ are the central charges of the Virasoro and Kac-Moody algebras respectively. Under the global symmetries~\eqref{E:WCFTsymmetries}, the stress tensor and momentum transform as~\cite{Detournay:2012pc}
\begin{align}
\begin{split}
\label{E:newTP}
\hat{T}(X) & = \left( \frac{\partial x^-}{\partial X^-}\right)^2 \left( T(x) - \frac{c}{12}\{  X^-(x^-),x^-\}\right) + \frac{\partial x^-}{\partial X^-}\frac{\partial x^+}{\partial X^-}P(x) - \frac{k}{4}\left( \frac{\partial x^+}{\partial X^-}\right)^2\,,
\\
\hat{P}(X) & = \frac{\partial x^-}{\partial X^-}P(x) - \frac{k}{2}\frac{\partial x^+}{\partial X^-}\,.
\end{split}
\end{align} 
where 
\beq
\{ f(x),x\} =\frac{f'''(x)}{f'(x)}-\frac{3}{2}\left(\frac{f''(x)}{f'(x)}\right)^2\,,
\eeq
is the Schwarzian derivative.

The authors of~\cite{Detournay:2012pc} used the symmetries~\eqref{E:WCFTsymmetries} and transformation laws~\eqref{E:newTP} to deduce a modular transformation law of the torus partition function under an ``$S$'' transformation which exchanges thermal and spatial circles. Under some assumptions regarding the vevs of the stress tensor and momentum current at zero temperature, those authors thereby obtained a warped analogue of the Cardy formula for $2d$ CFT.

We revisit WCFT thermodynamics later in Section~\ref{S:thermodynamics}. In particular, we show that infinite-volume WCFT thermodynamics immediately follows from a warped conformal map from the Euclidean plane to the thermal cylinder.

\subsection{Warped geometry}

In the same way that many useful properties of ordinary CFT, like its central charges, are visible when putting it on a fixed spacetime, it is also useful to couple WCFTs to the warped analogue of a metric.

While ordinary CFTs naturally couple to an external Riemannian metric, WCFTs naturally couple~\cite{Hofman:2014loa} to a structure known as ``Newton-Cartan'' (NC) geometry. There are several versions of NC geometry; some are relevant for Galilean field theory~\cite{Geracie:2014nka,Jensen:2014aia}, the Newtonian limit of General Relativity~\cite{Geracie:2015dea}, non-relativistic holography~\cite{Hartong:2015wxa}, or non-relativistic systems without a boost symmetry. The NC geometry relevant for WCFTs was worked out in~\cite{Hofman:2014loa} (see also~\cite{Hartong:2015xda,Hartong:2015usd}), where it was dubbed ``warped geometry.''

In this Subsection we review and rewrite warped geometry in a way which will be useful when we classify the anomalies that arise when a WCFT is coupled to a curved background. Our approach closely follows that of~\cite{Hartong:2015xda,Hartong:2015usd}.

Warped geometry is simple. Let $\M$ be an orientable manifold on which we will put a WCFT. We endow $\M$ with two nowhere-vanishing, linearly independent one-forms $n_{\mu}$ and $h_{\nu}$ subject to two local redundancies: a ``Weyl'' rescaling and a ``boost.'' Under these transformations, $n_{\mu}$ and $h_{\mu}$ transform as
\begin{align}
\begin{split}
\label{E:weylBoost}
\text{``Weyl''}:\,\,\, & n_{\mu} \to e^{\Omega}n_{\mu}\,, \qquad h_{\mu}\to h_{\mu}\,,
\\
\text{``Boost''}: \,\,\, & n_{\mu} \to n_{\mu}\,, \qquad \quad h_{\mu}\to h_{\mu} - \psi n_{\mu}\,,
\end{split}
\end{align}
where $\Omega$ and $\psi$ are functions which characterize the rescaling and boost. Roughly speaking, $n$ is the more general version of the ``right-moving'' direction $dx^-$, and $h$ the more general version of $dx^+$. 

From $n_{\mu}$ and $h_{\nu}$ we algebraically obtain two vector fields $v^{\mu}$ and $w^{\mu}$ satisfying
\beq
v^{\mu}n_{\mu}=w^{\mu}h_{\mu}=1\,, \qquad v^{\mu}h_{\mu}=w^{\mu}n_{\mu}=0\,.
\eeq
They transform under Weyl rescaling and boosts as
\begin{align}
\begin{split}
\text{``Weyl''}:&  \,\,\, v^{\mu} \to e^{-\Omega}v^{\mu}\,, \qquad \quad\,\, w^{\mu}\to w^{\mu}\,,
\\
\text{``Boost''}: &\,\,\, v^{\mu} \to v^{\mu} + \psi w^{\mu}\,, \qquad w^{\mu}\to w^{\mu}\,.
\end{split}
\end{align}
We also construct a boost-invariant measure $d^2x \sqrt{\gamma}$ with $\sqrt{\gamma}=\sqrt{\text{det}(\gamma_{\mu\nu})}$ and
\beq
\gamma_{\mu\nu}\equiv n_{\mu}n_{\nu}+h_{\mu}h_{\nu}\,,
\eeq
along with a boost-invariant epsilon tensor
\beq
\label{E:epsilon}
\varepsilon^{\mu\nu} = v^{\mu}w^{\nu}-v^{\nu}w^{\mu}=\frac{\epsilon^{\mu\nu}}{\sqrt{\gamma}}\,, \qquad \varepsilon_{\mu\nu} = n_{\mu}h_{\nu}-n_{\nu}h_{\mu}=\sqrt{\gamma}\,\epsilon_{\mu\nu}\,,
\eeq
where $\epsilon^{\mu\nu}$ and $\epsilon_{\mu\nu}$ are epsilon symbols. The measure transforms homogeneously as $\sqrt{\gamma}\to e^{\Omega}\sqrt{\gamma}$ under Weyl rescaling.

A few words are in order regarding the global structure of the geometry. We have in mind a geometry in which $(n_{\mu},h_{\nu})$ are globally-defined and everywhere non-vanishing, and that the WCFT in question is coupled to these fields in a Weyl and boost-invariant way. $\mathcal{M}$ must then have the topology of a cylinder or a torus. Perhaps one may define the geometry in such a way that $(n_{\mu},h_{\nu})$ are not globally-defined, but instead are sections of some bundle.

The relation to ordinary NC geometry is the following. The basic building blocks of NC geometry are the tensors $(n_{\mu},h_{\nu\rho})$, where $n_{\mu}$ is nowhere-vanishing, $h_{\mu\nu}$ is a spatial metric, and the two are constrained so that $n_{\mu}n_{\nu}+h_{\mu\nu}$ is positive-definite. In two spacetime dimensions, $h_{\mu\nu}$ can be expressed in terms of a one-form via $h_{\mu\nu}=h_{\mu}h_{\nu}$. The one-forms $n_{\mu}$ and $h_{\mu}$ are those described above.

As for the Weyl and boost symmetries, these also have counterparts in NC geometry. The NC geometry to which Galilean field theories couple possesses a boost symmetry. This transformation, known as a Milne boost, leaves $n_{\mu}$ invariant but shifts $h_{\mu\nu}$ as
\beq
\label{E:NCmilne}
h_{\mu\nu} \to h_{\mu\nu} - \left( n_{\mu}h_{\nu\rho} + n_{\nu}h_{\mu\rho}\right)\psi^{\rho} + n_{\mu}n_{\nu}h_{\rho\sigma}\psi^{\rho}\psi^{\sigma}\,, \qquad \psi^{\mu}n_{\mu} = 0\,.
\eeq
The Milne symmetry is intimately related to Galilean invariance as described in~\cite{Jensen:2014aia}. In two spacetime dimensions where $h_{\mu\nu}=h_{\mu}h_{\nu}$, the Milne boost~\eqref{E:NCmilne} becomes the boost in~\eqref{E:weylBoost} with $\psi = h_{\mu}\psi^{\mu}$. For Weyl, recall that non-relativistic scale-invariant theories are characterized by a dynamical exponent $z$. When such a theory is coupled to NC geometry, the theory is invariant under local ``Weyl'' rescalings of $n_{\mu}$ and $h_{\nu\rho}$ as
\beq
n_{\mu} \to e^{ \Omega} n_{\mu}\,, \qquad h_{\mu\nu} \to e^{2\Omega/z}h_{\mu\nu}\,.
\eeq
In two spacetime dimensions with $z=\infty$, this becomes the Weyl rescaling in~\eqref{E:weylBoost}.

From the one-forms $(n_{\mu},h_{\nu})$ we define a covariant derivative $D_{\mu}$ via the connection
\beq
\label{E:Gamma}
\Gamma^{\mu}{}_{\nu\rho} = v^{\mu}\partial_{\rho}n_{\nu} + w^{\mu}\partial_{\rho}h_{\nu}\,.
\eeq
This connection is neither boost nor Weyl invariant. Under an infinitesimal reparameterization $\xi^{\mu}$, Weyl rescaling $\omega$, and boost $\psi$, which we colletively denote as $\chi = (\xi^{\mu},\omega,\psi)$, the connection one-form $\Gamma^{\mu}{}_{\nu} = \Gamma^{\mu}{}_{\nu\rho} dx^{\rho}$ varies as
\beq
\delta_{\chi}\Gamma^{\mu}{}_{\nu}=\pounds_{\xi}\Gamma^{\mu}{}_{\nu} + d\partial_{\nu}\xi^{\mu} + v^{\mu}n_{\nu}d\omega - w^{\mu}n_{\nu}d\psi\,,
\eeq
where in the first term we define the action of the Lie derivative $\pounds_{\xi}$ along $\xi^{\mu}$ as though $\Gamma^{\mu}{}_{\nu}$ were a tensor. The derivative $D_{\mu}$ has nonzero torsion but zero curvature,
\begin{align}
\begin{split}
\label{E:RT}
R^{\mu}{}_{\nu} &= d\Gamma^{\mu}{}_{\nu} + \Gamma^{\mu}{}_{\rho}\wedge \Gamma^{\rho}{}_{\nu} = \frac{1}{2}R^{\mu}{}_{\nu\rho\sigma}dx^{\rho}\wedge dx^{\sigma} = 0\,,
\\
T^{\mu} & = \frac{1}{2}T^{\mu}{}_{\nu\rho}dx^{\nu}\wedge dx^{\rho} = - \frac{1}{2}\Gamma^{\mu}{}_{\nu\rho}dx^{\nu}\wedge dx^{\rho} = -v^{\mu}dn - w^{\mu}dh\,.
\end{split}
\end{align}
The torsion is determined in terms of the two scalars
\beq
\label{E:defineNH}
N \equiv \varepsilon^{\mu\nu}\partial_{\mu}n_{\nu}\,, \qquad H \equiv \varepsilon^{\mu\nu}\partial_{\mu}h_{\nu}\,.
\eeq
The derivative also fixes $(n_{\mu},h_{\nu})$,
\beq
D_{\mu}n_{\nu} = D_{\mu}h_{\nu} = 0\,.
\eeq

There is a first-order version of this geometry. Because $(n_{\mu},h_{\nu})$ are globally defined, the frame bundle is topologically trivial. We trivialize it by restricting the frame to be $F_A^{\mu} = (v^{\mu},w^{\nu})$. There is no spin connection, and~\eqref{E:RT} follows.

There is another first-order formulation of NC geometry which naturally appears in Lifshitz holography (see e.g.~\cite{Christensen:2013rfa,Hartong:2015wxa}). Given $n_{\mu}$ and the inverse spatial metric $h^{\mu\nu}$, one takes $n_{\mu}$ to be the temporal component of the coframe and decomposes $h^{\mu\nu}$ into inverse spatial vielbeins. This decomposition is invariant under local spatial rotations as well as under local boosts. Fixing the temporal component of the frame to be $v^{\mu}$, these boosts act in the same way as a Milne boost. In this way, the local boost symmetry is ``absorbed'' into the local redefinitions of frame.

Another form of NC geometry was described in~\cite{Geracie:2015dea}, which naturally arises in the Newtonian limit of General Relativity. Its $d=2$ version (which of course has no relation to General Relativity) consists of $(n_{\mu},h_{\nu},\Omega,T^{\rho})$ where $T^{\mu}$ is the (dualized) torsion satisfying $n_{\mu}T^{\mu} = -N$ , and $\Omega$ is a scalar which transforms under infinitesimal boosts as
\beq
\delta_{\psi} \Omega = \varepsilon^{\mu\nu}\partial_{\mu}(\psi h_{\nu}) - \psi h_{\mu}T^{\mu}\,.
\eeq
One can use $\Omega$ to obtain a boost-invariant connection (although not a Weyl-invariant one). This geometry is interesting and useful, but is unrelated to WCFT for reasons we explain in the next Subsection.

\subsection{WCFTs are Galilean CFTs with $z=\infty$}
\label{S:galilean}

As WCFTs couple to the same geometry and possess the same invariances as Galilean CFTs with dynamical critical exponent $z=\infty$,\footnote{Here we mean Galilean theories with no particle number symmetry.} we may as well identify the two.

Before going on, let us argue that the ``warped geometry'' above is the correct one to which WCFTs couple. We do so by showing that the transformations~\eqref{E:WCFTsymmetries} are exactly the ``warped isometries'' of the flat structure $n=dx^-, h=dx^+$, and therefore correspond to global symmetries of the flat-space WCFT. At the infinitesimal level, we need the combination of a coordinate reparameterization $\xi^{\mu}$, Weyl rescaling $\sigma$, and boost $\psi$ that fix the flat background. Collectively notating the transformation as $\chi = (\xi^{\mu},\sigma,\psi)$ and its action as $\delta_{\chi}$, we have
\begin{align}
\begin{split}
\delta_{\chi}n_{\mu} & = \pounds_{\xi} n_{\mu} + \sigma n_{\mu}\,,
\\
\delta_{\chi}h_{\mu} & = \pounds_{\xi}h_{\mu} - \psi n_{\mu}\,,
\end{split}
\end{align}
with $\pounds_{\xi}$ the Lie derivative along $\xi^{\mu}$. Setting $\delta_{\chi}n_{\mu}=\delta_{\chi}h_{\mu}=0$ we obtain the warped isometries. They are parameterized by two free functions of $x^-$ as
\beq
\xi^{\mu}\partial_{\mu} = f_1(x^-)\partial_- + f_2(x^-)\partial_+\,, \qquad \sigma = - f_1'(x^-)\,, \qquad \psi = f_2'(x^-)\,,
\eeq
which exponentiate to
\beq
\label{E:WCFTsymmetries2}
x^- = x^-(X^-)\,, \qquad x^+ = X^++f(X^-)\,, \qquad e^{\Omega} = \frac{\partial X^-}{\partial x^-}\,, \qquad \psi = f'(X^-)\,.
\eeq
As claimed, the coordinate transformations are the WCFT symmetries~\eqref{E:WCFTsymmetries}.

These transformations include ordinary Galilean boosts~\cite{Hofman:2014loa}. To guide the eye, set $x^- = t, x^+ = x$. The symmetry
\beq
t =T\,, \qquad x = X - v T\,,
\eeq
is just a Galilean boost. This also suggests that we ought to identify $x^-$ with time and $x^+$ with space.

We record five observations before proceeding.
\begin{enumerate}
\item The infinite-dimensional symmetry is not specific to $z=\infty$. It is easy to show that a Galilean theory with generic $z$ has a flat-space symmetry group parameterized by two free functions of $x^-$.
\item Warped geometry has the analogue of ``conformal gauge:'' one can use the combination of a coordinate transformation, Weyl rescaling, and boost, to locally set $n=dx^-, h=dx^+$.
\item WCFTs do not couple to the version of NC geometry appearing in~\cite{Geracie:2015dea} we mentioned in the previous Subsection; the ``warped isometries'' of the flat structure of that version ($n=dx^-,h=dx^+,\Omega=0,T^{\mu}=0$) are not~\eqref{E:WCFTsymmetries}. For example, $\Omega$ is not invariant under the coordinate transformation $x^+=X^+ + f(X^-)$ and boost $\psi = f'(X^-)$ which fix $(n_{\mu},h_{\nu},T^{\rho})$.
\item In two dimensions, there is an isomorphism between the (massless) Galilean algebra and the Carroll algebra. (By the massless Galilean algebra we mean the Galilean algebra without a central extension, and so without a conserved particle number.) The Carroll algebra arises in the $c\to 0$ limit of relativistic systems. So WCFTs are also Carrollian CFTs, as pointed out in~\cite{Hartong:2015xda}. Relatedly~\cite{Hartong:2015xda}, the warped geometry above is the ``Carrollian geometry'' to which Carrollian theories naturally couple. This point of view is useful as the $d>2$ version of warped theories suggested by~\cite{Hofman:2014loa} are Carrollian.
\item Later, in Section~\ref{S:holography} when we study gravity on so-called warped AdS$_3$ spacetimes, that the boundary is equipped with a warped geometry subject to a constraint. The one-form $n$ is obeys $dn=0$, and the allowed Weyl transformations respect $w^{\mu}\partial_{\mu}\Omega=0$.
\end{enumerate}

\section{Free WCFT}
\label{S:freeWCFT}

Using their ``warped geometry,'' Hofman and Rollier~\cite{Hofman:2014loa} have recently obtained the first Lagrangian examples of WCFTs. They found two free theories: (i.) a Weyl fermion with a Lorentz-violating mass, and (ii.) a warped analogue of a $bc$ system. We review their theories, and then go on to construct two different free scalar WCFTs. We find that all of these WCFTs are perched on the edge of non-locality. 

A word about warped dimensional analysis is in order. The warped scaling acts as $x^-\to \lambda x^-$, $x^+\to x^+$. So we assign $x^-$ dimension $-1$ and $x^+$ dimension $0$, and a WCFT Lagrangian ought to carry dimension $1$. Irrelevant operators have dimension greater than $1$, relevant less than $1$, and exactly marginal carry dimension $1$.

\subsection{The Hofman/Rollier theories}

Let us summarize the free theories constructed in~\cite{Hofman:2014loa} in the language of this paper. The basic field is a two-component anticommuting field $\Psi = (\Psi^-,\Psi^+)$ which transforms under local boosts as
\beq
\Psi^+ \to \Psi^+ + \frac{\psi}{2}\Psi^-\,, \qquad \Psi^- \to \Psi^-\,.
\eeq
The conjugate field $\overline{\Psi} = (\overline{\Psi}^-,\overline{\Psi}^+)$ transforms in the same way,
\beq
\overline{\Psi}^+ \to \overline{\Psi}^+ + \frac{\psi}{2}\overline{\Psi}^-\,, \qquad \overline{\Psi}^-\to \overline{\Psi}^-\,.
\eeq

For the warped Weyl theory we let $\Psi$ be a complex spinor. The action is
\beq
\label{E:Sweyl}
S_{weyl} = \int d^2x \sqrt{\gamma} \Big\{ i \,\overline{\Psi}^- w^{\mu}\partial_{\mu}\Psi^- + m \overline{\Psi}^- \Psi^-\Big\}\,.
\eeq
This theory is manifestly boost-invariant. It is also Weyl-invariant if we assign the transformation law
\beq
\Psi^- \to e^{-\frac{\Omega}{2}}\Psi^-\,, \qquad \overline{\Psi}^- \to e^{-\frac{\Omega}{2}}\overline{\Psi}^-\,,
\eeq
under Weyl rescalings, i.e. the fermion fields carry a free-field dimension of $1/2$. 

For the warped $bc$ theory we let $\Psi$ be a real spinor. The action is
\beq
\label{E:Sbc}
S_{bc} = \int d^2x \sqrt{\gamma} \Big\{ i \, \Psi^- v^{\mu}\partial_{\mu} \Psi^- - 2i \, \Psi^+ w^{\mu}\partial_{\mu}\Psi^- - 2 m \Psi^+ \Psi^-\Big\}\,,
\eeq
This action is boost-invariant: the boost variation of the first term is $i \psi \Psi^- w^{\mu}\partial_{\mu} \Psi^-$, which offsets the variation of the second term upon using that the boost variation of $\Psi^+$ is $\frac{\psi}{2}\Psi^-$. The boost variation of the third term vanishes on account of $(\Psi^-)^2=0$. It is also Weyl-invariant if we assign
\beq
\Psi^- \to \Psi^-\,, \qquad \Psi^+ \to e^{-\Omega} \Psi^+\,,
\eeq
under Weyl rescalings. Notice that $\Psi^+$ is a Legendre multiplier field which enforces the constraint
\beq
\label{E:bcConstraint}
w^{\mu}\partial_{\mu}\Psi^- - i m \Psi^- = 0\,.
\eeq

In flat space ($n=dx^-\,, h = dx^+$) these theories are simply
\begin{align}
\begin{split}
S_{weyl} & = \int d^2x \, \overline{\Psi}^- i(\partial_+ - i m ) \Psi^-\,,
\\
S_{bc} &= \int d^2x \left(  \Psi^- \, i \partial_-\Psi^- - 2i \Psi^+ (\partial_+\Psi^- - i m \Psi^-)\right) \,,
\end{split}
\end{align}
where in the warped $bc$ system $\Psi^-$ is a real anticommuting field subject to the constraint~\eqref{E:bcConstraint}. We see that the warped Weyl theory is essentially the theory of an ordinary Weyl fermion $\chi$ at chemical potential $m$. Indeed, the momentum current $m \overline{\Psi}^-\Psi^-$ is proportional to the $U(1)$ current $\bar{\chi}\chi$ upon identifying $\Psi^- \to \chi$.

Now for the punchline of this Subsection. Owing to the fact that $w^{\mu}\partial_{\mu}$ is a boost and Weyl-invariant differential operator, both the warped Weyl fermion and warped $bc$ system admit an infinite number of exactly marginal deformations. For the Weyl fermion, the operator $i\, \overline{\Psi}^- (\partial_+)^n\Psi^- $ is boost-invariant and exactly marginal for any $n$. For the $bc$ system, the operator that does the job is $i \, \overline{\Psi}^- \partial_- (\partial_+)^n\Psi^- - 2 i \overline{\Psi}^+ (\partial_+)^{n+1}\Psi^-$. Since there are exactly marginal deformations with an arbitrarily large number of $+$ derivatives, the set of these operators may be reexpressed in terms of an infinite number of exactly marginal operators which are non-local along $x^+$. For the Weyl fermion the operators above are equivalent to $\overline{\Psi}^-(x) \exp\left( \ell \partial_+\right) \Psi^-(x)$ which is clearly non-local as well as exactly marginal. We then see that both of the Hofman/Rollier theories are on the brink of being non-local in the $x^+$ direction: to obtain them, we must tune away this infinite family of operators.\footnote{It may be that global constraints on a WCFT, like those studied in~\cite{Castro:2015uaa}, would disallow these non-local deformations here, or for the scalar WCFTs we write down below.}

Even having tuned away these non-local operators, the Hofman/Rollier WCFTs have a paucity of relevant deformations. A theory with multiple species of Weyl fermions or several $bc$ systems will generically have some flavor symmetry. However that flavor symmetry is always anomalous and so cannot be gauged. Using that the Weyl fermions carry dimension $1/2$ while the $bc$ fermions (subject to constraints~\eqref{E:bcConstraint}) are dimensionless, we learn that the classically relevant operators are polynomials in at least two species of $bc$ fermions. There are also classically marginal operators built from those polynomials times Weyl bilinears.

In the remainder of this Section we obtain scalar analogues of the Hofman/Rollier theories. They satisfy the same essential properties we found above: (i.) they too are on the precipice of non-locality in the $+$ direction, and (ii.) have a similar classification of relevant and classically marginal operators.

\subsection{Scalars, take one}

Let us try to write down a WCFT of a free real scalar $\varphi$ in the same vein as the warped Weyl theory~\eqref{E:Sweyl}. It is likely non-unitary.

Using that $w^{\mu}\partial_{\mu}$ is a boost and Weyl-invariant differential operator, we could construct a theory whose kinetic term is $(w^{\mu}\partial_{\mu}\varphi)^2$. The unique two-derivative action with this kinetic term, consistent with the warped symmetries, is
\beq
\label{E:Sfree}
S_1= \int d^2x \sqrt{\gamma} \left\{ \frac{1}{2}(w^{\mu}\partial_{\mu}\varphi)^2  + \frac{1}{4}\left(w^{\mu} \partial_{\mu}N - \frac{N^2}{2}\right) \varphi^2 - \frac{m^2}{2} \varphi^2 \right\}\,
\eeq
Recall $N = \varepsilon^{\mu\nu}\partial_{\mu}n_{\nu}$. This action is Weyl and boost-invariant provided that $\varphi$ is boost-invariant and transforms as
\beq
\varphi \to e^{-\frac{\Omega}{2}}\varphi\,,
\eeq
under Weyl rescalings, i.e. $\varphi$ carries a free-field dimension of $1/2$. The first term in~\eqref{E:Sfree} is the scalar kinetic term, the second is the warped analogue of the conformal coupling to curvature, and the third, despite looking like a mass term, is allowed by the symmetries of the problem as $\sqrt{\gamma}\varphi^2$ is Weyl-invariant. In flat space, this action is
\beq
S_1 = \int d^2x \left\{ \frac{1}{2}(\partial_+\varphi)^2 - \frac{m^2}{2}\varphi^2\right\}\,.
\eeq
The field equation for $\varphi$ is solved by
\beq
\varphi = \varphi_+(x^-)e^{im x^+} + \varphi_-(x^-) e^{-im x^+}\,,
\eeq
After Wick-rotating $x^-$ to obtain a Euclidean version of this theory, the Euclidean action~\eqref{E:Sfree} is not bounded below and so we expect that this theory is non-unitary.

The stress tensor $T$ and momentum $P$ are given by
\beq
T = \frac{1}{4\pi}\left( \partial_- \varphi\partial_+ \varphi - \varphi \partial_-\partial_+\varphi\right)\,, \qquad P = \frac{1}{4\pi}\left( (\partial_+\varphi)^2 - m^2 \varphi^2\right)\,,
\eeq
satisfying
\beq
\partial_+ T = 0\,, \qquad \partial_+ P = 0\,,
\eeq
on-shell.

Given a complex scalar $\Phi$ we can construct an even simpler WCFT,
\beq
S_2 = \int d^2x \left\{ \frac{i}{2}w^{\mu}\Big( \Phi^{\dagger}\partial_{\mu}\Phi - (\partial_{\mu}\Phi^{\dagger})\Phi\Big) + m |\Phi|^2\right\}\,,
\eeq
which is invariant under the warped symmetries if $\Phi$ is boost-invariant and carries dimension $1/2$. Observe that this theory has a $U(1)$ flavor symmetry and that the ``mass'' term $m|\Phi|^2$ can be absorbed into a background gauge field $A_{\mu}$ which couples to the flavor symmetry satisfying $w^{\mu}A_{\mu} = m$. The parameter $m$ sets the Kac-Moody level.

These scalars share two features with the warped Weyl theory: (i.) they admit an infinite number of exactly marginal non-local deformations, and (ii.) they possess no classically relevant operators.

\subsection{Scalars, take two}
\label{S:scalar2}

Now we build the scalar analogue of the warped $bc$ theory~\eqref{E:Sbc}. Let $\varphi$ and $\eta$ be real scalars. Consider the action
\beq
\label{E:Sfree2}
S_3 = \int d^2x \sqrt{\gamma} \Big\{ (v^{\mu}\partial_{\mu}\varphi)(w^{\nu}\partial_{\nu}\varphi) +\eta w^{\mu}\partial_{\mu}\varphi- m H \varphi\Big\}\,,
\eeq
where we remind the reader that $H = \varepsilon^{\mu\nu}\partial_{\mu}h_{\nu}$. The last term can be written after an integration by parts as $  m v^{\mu}\partial_{\mu}\varphi$. Note that $\eta$ is a Lagrange multiplier field enforcing the constraint
\beq
w^{\mu}\partial_{\mu}\varphi = 0\,.
\eeq
The action $S_3$ is invariant under the warped symmetries upon assigning $\eta$ and $\varphi$ the boost transformations
\beq
\eta \to \eta + \psi\, m - \psi\,w^{\mu}\partial_{\mu}\varphi\,, \qquad \varphi \to \varphi
\eeq
and dimensions $1$ and $0$ respectively. In flat space the action becomes
\beq
S_3 = \int d^2x \Big\{ \partial_- \varphi \partial_+ \varphi  + \eta \partial_+ \varphi \Big\}\,.
\eeq
The equations of motion imply that $\varphi$ and $\eta$ are chiral,
\beq
\varphi = \varphi(x^-)\,, \qquad \eta = \eta(x^-)\,.
\eeq

This theory is essentially that of two ordinary right-moving scalars, with $\varphi$ linearly coupled to an external gauge field $A_{\mu}$ under the substitution $A_{\mu} \to h_{\mu}$. Indeed, the stress tensor and momentum current
\beq
T  = \frac{(\partial_-\varphi)^2+\eta \partial_-\varphi}{2\pi}\,, \qquad P  = \frac{m\partial_-\varphi}{2\pi}\,,
\eeq
are (ignoring the term coming from $\eta$) the stress tensor and conserved current of the real chiral scalar. 

This scalar WCFT shares the same qualitative features as the ones we pointed out for the warped $bc$ theory: (i.) it admits an infinite number of exactly marginal non-local deformations, and (ii.) by dimensional analysis the only classically relevant operators are polynomials in $\varphi$.

\section{Anomalies}
\label{S:anomalies}

In the previous Section we saw that Lagrangian WCFTs require an infinite fine-tuning to be rendered local. Now we turn to a non-perturbative analysis of local WCFTs via their `t Hooft anomalies.\footnote{There has been some prior work on anomalies in Galilean field theories, e.g.~\cite{Jensen:2014hqa,Arav:2016xjc,Pal:2017ntk,Auzzi:2017jry,Fernandes:2017nvx}. The anomalies uncovered in that work however have zero intersection with those found here.} The goal of this Section is three-fold. First, to obtain the anomalies of local WCFTs, then to derive the anomalous Ward identities, and finally to relate the anomaly coefficients to the Virasoro and Kac-Moody central charges. Before doing so, we warn the reader that this Section is a bit technical. The anomalies are presented in~\eqref{E:anomalies}, the Ward identities in~\eqref{E:anomWard}, and the relation to central charges in~\eqref{E:tildecToc}.

Throughout, our conventions are that the generating functional of connected, real-time correlation functions $W$ is given by
\beq
W = - i \ln \mathcal{Z}\,,
\eeq
with $\mathcal{Z}$ the WCFT partition function, and that connected one-point functions of the WCFT ``stress tensor'' are obtained by variation of $W$ with respect to the background geometry $(n_{\mu},h_{\nu})$,
\beq
\label{E:deltaW}
\delta W = -\frac{1}{2\pi}\int d^2x \sqrt{\gamma} \Big\{ \mathcal{N}^{\mu}\delta n_{\mu} + \mathcal{H}^{\mu}\delta h_{\mu}\Big\}\,.
\eeq

Our classification is entirely ``local,'' insofar as we consider infinitesimal symmetry transformations. We ignore ``global'' constraints like the modular constraints on the torus partition function considered in~\cite{Castro:2015uaa}.

\subsection{The classification}

Anomalies are strongly constrained by Wess-Zumino (WZ) consistency. Under an infinitesimal symmetry transformation $\delta_{\chi}$, we require the integrability condition
\beq
\label{E:algebra}
[\delta_{\chi_1},\delta_{\chi_2}] -\delta_{\chi_{[12]}}=0\,,
\eeq
that is, the symmetry transformations generate an algebra. What is this algebra? As we mentioned above, we have
\beq
\label{E:deltaChiNC}
\delta_{\chi}n_{\mu}  = \pounds_{\xi}n_{\mu} + \sigma \, n_{\mu}\,, \qquad 
\delta_{\chi}h_{\mu}  = \pounds_{\xi}h_{\mu} - \psi \, n_{\mu}\,.
\eeq
In order for $n_{\mu} + \delta_{\chi}n_{\mu}$ and $h_{\nu}+\delta_{\chi}h_{\nu}$ to be tensors which transform in the same way as $n_{\mu}$ and $h_{\nu}$, e.g.
\beq
\delta_{\chi_1}\big( \delta_{\chi_2}n_{\mu}\big) = \pounds_{\xi_1} \big( \delta_{\chi_2}n_{\mu}\big) + \sigma_1 \delta_{\chi_2}n_{\mu}\,,
\eeq
the transformation parameters must vary as
\begin{align}
\begin{split}
\label{E:deltaSymmetry}
\delta_{\chi_1}\xi_2^{\mu} &= \pounds_{\xi_1}\xi_2^{\mu} = \xi_1^{\nu}\partial_{\nu}\xi_2^{\mu}-\xi_2^{\nu}\partial_{\nu}\xi_1^{\mu}\,,
\\
\delta_{\chi_1}\psi_2 & = \pounds_{\xi_1}\psi_2- \pounds_{\xi_2}\psi_1 - \sigma_1 \, \psi_2 + \sigma_2 \,\psi_1=\xi_1^{\mu}\partial_{\mu}\psi_2-\xi_2^{\mu}\partial_{\mu}\psi_1 -\sigma_1\, \psi_2 + \sigma_2\, \psi_1\,, 
\\
\delta_{\chi_1}\sigma_2 & =\pounds_{\xi_1} \sigma_2 - \pounds_{\xi_2} \sigma_1 = \xi_1^{\mu}\partial_{\mu}\sigma_2 - \xi_2^{\mu}\partial_{\mu}\sigma_1\,.
\end{split}
\end{align}
Then~\eqref{E:algebra} holds with $\chi_{[12]} = (\xi_{[12]}^{\mu},\psi_{[12]},\sigma_{[12]})$ and
\beq
\label{E:[12]}
\xi_{[12]}^{\mu}  = \pounds_{\xi_1}\xi_2^{\mu}\,, \qquad \psi_{[12]} = \pounds_{\xi_1}\psi_2 - \pounds_{\xi_2}\psi_1 - \sigma_1 \,\psi_2 + \sigma_2 \,\psi_1\,, \qquad \sigma_{[12]} = \pounds_{\xi_1}\sigma_2 - \pounds_{\xi_2}\sigma_1\,.
\eeq

In a non-anomalous theory, the generating functional $W$ is invariant under infinitesimal coordinate reparameterizations, \&c, meaning $\delta_{\chi}W=0$. In an anomalous theory,  $\delta_{\chi}W \neq 0$ and the anomalies are specified by this variation. We parameterize
\beq
\label{E:deltaChiW}
\delta_{\chi}W =\frac{1}{2\pi} \int d^2x \sqrt{\gamma} \left\{\partial_{\mu} \xi^{\nu} \mathcal{T}^{\mu}{}_{\nu} + \psi \mathcal{P} + \Omega \mathcal{A}\right\}\,.
\eeq
where $\mathcal{T}^{\mu}{}_{\nu}, \mathcal{P},$ and $\mathcal{A}$ are built from the background fields.

We use a three-step algorithm to classify anomalies: (i.) we parameterize the most general anomalies, (ii.) use the most general local counterterms to remove terms in $\delta_{\chi}W$, and (iii.) impose WZ consistency~\eqref{E:algebra}.

Anomalies are sensitive to locality precisely because of this second step, the modding out by local counterterms. After all, anomalies may always be removed by the addition of non-local counterterms.

As a practical matter, we perform this algorithm order by order in gradients up to two derivative terms in $(\mathcal{T}^{\mu}{}_{\nu},\mathcal{P},\mathcal{A})$. In a few places we will use the shorthand notation
\beq
\dot{\mathcal{X}} = v^{\mu}\partial_{\mu}\mathcal{X}\,, \qquad \mathcal{X}' = w^{\mu}\partial_{\mu}\mathcal{X}\,.
\eeq

This algorithm is straightforward to employ and leads to three anomalies parameterized by three anomaly coefficients $(\tilde{k},\tilde{c}_1,\tilde{c}_2)$. The first is a pure boost anomaly,
\begin{subequations}
\label{E:anomalies}
\beq
\label{E:boostAnomaly}
\delta_{\chi}W_{\tilde{k}} = \frac{\tilde{k}}{8\pi}\int d^2x \sqrt{\gamma}\,\psi\,,
\eeq
the second is a mixed boost/Weyl anomaly,
\beq
\label{E:mixedAnomaly}
\delta_{\chi}W_1 = \frac{\tilde{c}_1}{24\pi}\int d^2x \sqrt{\gamma} \left( - \dot{\sigma}N + \frac{\psi}{2}N^2\right)\,,
\eeq
and the third is a gravitational anomaly (accompanied by boost and Weyl variations so as to be WZ consistent)
\beq
\label{E:gravAnomaly}
\delta_{\chi}W_2  = - \frac{\tilde{c}_2}{192\pi} \int \left( \partial_{\mu}\xi^{\nu}d\Gamma^{\mu}{}_{\nu} - d\psi \wedge \Gamma^{\mu}{}_{\nu}w^{\nu}n_{\mu} + d\sigma \wedge \Gamma^{\mu}{}_{\nu}v^{\nu}n_{\mu}\right)\,.
\eeq
\end{subequations}
Later we will see that $\tilde{k}$ and $\tilde{c}_1$ respectively determine the Kac-Moody central charge $k$ and Virasoro central charge $c$ of the underlying WCFT.

In many ways, the mixed boost/Weyl anomaly is like the ordinary Weyl anomaly of $2d$ CFT, and the gravitational anomaly like that of the gravitational anomaly of $2d$ CFT. 

Let us show that the anomalies are consistent. We begin with the boost anomaly~\eqref{E:boostAnomaly}, whose second variation is
\beq
\delta_{\chi_1}\delta_{\chi_2}W_{\tilde{k}} = \frac{\tilde{k}}{8\pi}\int d^2x\sqrt{\gamma} \big( - \pounds_{\xi_2}\psi_1 + \sigma_2 \psi_1\big)\,.
\eeq
The commutator of variations is then
\begin{align}
\nonumber
[\delta_{\chi_1},\delta_{\chi_2}]W_{\tilde{k}}& = \frac{\tilde{k}}{8\pi}\int d^2x \sqrt{\gamma} \big( \pounds_{\xi_1}\psi_2 - \pounds_{\xi_2}\psi_1 - \sigma_1 \psi_2 + \sigma_2 \psi_1\big) = \frac{\tilde{k}}{8\pi}\int d^2x\sqrt{\gamma} \,\psi_{[12]}
\\
& = \delta_{\chi_{[12]}}W_{\tilde{k}}\,,
\end{align}
where we have used the algebra~\eqref{E:[12]} to go from the first line to the second. So the boost anomaly is consistent.

Now for the mixed boost/Weyl anomaly~\eqref{E:mixedAnomaly}. Its second variation is
\begin{align}
\begin{split}
\delta_{\chi_1}\delta_{\chi_2}W_1 =& \frac{\tilde{c}_1}{24\pi}\int d^2x \sqrt{\gamma} \Big\{ \dot{\sigma}_2\sigma_1'  + v^{\mu}\partial_{\mu}\left( \pounds_{\xi_2}\sigma_1\right) N+\left( - \pounds_{\xi_2}\psi_1 + \sigma_2 \psi_1 \right) \frac{N^2}{2} 
\\
& \qquad \qquad \qquad \qquad - \left( \psi_1 \sigma_2' +\psi_2 \sigma_1'\right)N \Big\}\,.
\end{split}
\end{align}
The commutator is
\begin{align}
\begin{split}
[\delta_{\chi_1},\delta_{\chi_2}]W_1 = &\frac{\tilde{c}_1}{24\pi}\int d^2x \sqrt{\gamma} \Big\{ \dot{\sigma}_2 \sigma_1' - \sigma_2' \dot{\sigma}_1 - v^{\mu}\partial_{\mu} \left( \pounds_{\xi_1}\sigma_2 - \pounds_{\xi_2}\sigma_1\right) N\,,
\\
& \qquad \qquad \qquad \qquad + \left( \pounds_{\xi_1}\psi_2 -\pounds_{\xi_2}\psi_1 - \sigma_1 \psi_2 + \sigma_2 \psi_1\right)\frac{N^2}{2}\Big\}\,
\\
= & \frac{\tilde{c}_1}{24\pi}\int d^2x \sqrt{\gamma} \left\{ -v^{\mu}\partial_{\mu}\left( \sigma_{[12]}\right) N + \frac{\psi_{[12]}}{2}N^2 \right\} 
\\
= & \delta_{\chi_{[12]}}W_1\,.
\end{split}
\end{align}
In going from the first equality to the second we have used that
\beq
\dot{\sigma}_2\sigma_1' - \sigma_2' \dot{\sigma}_1 = \varepsilon^{\mu\nu}\partial_{\mu}\sigma_2\partial_{\nu}\sigma_1\,.
\eeq
Observe that the Weyl variation is not itself Weyl-invariant; in this sense this anomaly is akin to the A-type anomalies of ordinary even-dimensional CFT.

It is amusing to see exactly how the gravitational anomaly obeys WZ consistency. To do so, we require the infinitesimal variation of the non-tensorial objects,
\begin{align}
\begin{split}
\label{E:deltaGamma}
\delta_{\chi} \Gamma^{\mu}{}_{\nu\rho} &= \pounds_{\xi} \Gamma^{\mu}{}_{\nu\rho} + \partial_{\nu}\partial_{\rho}\xi^{\mu} - w^{\mu}n_{\nu}\partial_{\rho}\psi + v^{\mu}n_{\nu}\partial_{\rho} \sigma\,,
\\
\delta_{\xi_1}\partial_{\rho}\partial_{\nu}\xi_2^{\mu}& =\partial_{\rho}\partial_{\nu}\xi_{[12]}^{\mu} =\pounds_{\xi_1}\partial_{\rho}\partial_{\nu}\xi_2^{\mu} - \pounds_{\xi_2}\partial_{\rho}\partial_{\nu}\xi_1^{\mu} = - \delta_{\xi_2}\partial_{\rho}\partial_{\nu}\xi_1^{\mu}\,.
\end{split}
\end{align}
Up to boundary terms, the second variation is
\begin{align}
\nonumber
\delta_{\chi_1}\delta_{\chi_2}W_2 = & -\frac{\tilde{c}_2}{192\pi}\Big\{\int   \pounds_{\xi_2}d\partial_{\mu}\xi_1^{\nu} \wedge\Gamma^{\mu}{}_{\nu}- d\left( - \pounds_{\xi_2}\psi_1  + \sigma_2 \psi_1\right) \wedge \Gamma^{\mu}{}_{\nu}w^{\nu}n_{\mu} - d\pounds_{\xi_2}\sigma_1 \wedge \Gamma^{\mu}{}_{\nu}v^{\nu}n_{\mu}
\\
&\qquad \qquad +\left\{  d\partial_{\mu}\xi_2^{\nu}\wedge \left( w^{\mu}d\psi_1- v^{\mu}d\sigma_1\right) + d\partial_{\mu}\xi_1^{\nu} \wedge \left( w^{\mu}d\psi_2 - v^{\mu}d\sigma_2\right)\right\} n_{\nu}
\\
\nonumber
&\qquad \qquad \qquad + \left( \psi_2 d\sigma_1 + \psi_1 d\sigma_2\right) \wedge \Gamma^{\mu}{}_{\nu}w^{\nu}n_{\mu} \Big\}\,.
\end{align}
The second and third lines are symmetric under exchange of $\chi_1$ and $\chi_2$ and so up to a boundary term we find
\begin{align}
\begin{split}
[\delta_{\chi_1},\delta_{\chi_2}]W_2 & =- \frac{\tilde{c}_2}{192\pi}\Big\{ \int \left(- \pounds_{\xi_1}d\partial_{\mu}\xi_2^{\nu} + \pounds_{\xi_2}d\partial_{\mu}\xi_1^{\nu}\right)\wedge \Gamma^{\mu}{}_{\nu}
\\
& \qquad \qquad \qquad  -d \left( \pounds_{\xi_1}\psi_2 - \pounds_{\xi_2}\psi_1 - \sigma_1\psi_2 + \sigma_2 \psi_1\right)\wedge \Gamma^{\mu}{}_{\nu}w^{\nu}n_{\mu} 
\\
& \qquad \qquad \qquad \qquad + d \left( \pounds_{\xi_1}\sigma_2 - \pounds_{\xi_2}\sigma_1\right) \wedge \Gamma^{\mu}{}_{\nu}v^{\nu}n_{\mu}\Big\}
\\
& = - \frac{\tilde{c}_2}{192\pi}\int \left\{ \partial_{\mu}\xi_{[12]}^{\nu}\wedge d\Gamma^{\mu}{}_{\nu} - d\psi_{[12]}\wedge \Gamma^{\mu}{}_{\nu}w^{\nu}n_{\mu} + d\sigma_{[12]}\wedge \Gamma^{\mu}{}_{\nu}v^{\nu}n_{\mu}\right\}
\\
& = \delta_{\chi_{[12]}}W_2\,.
\end{split}
\end{align}

We conclude this Subsection with a few comments on the gravitational anomaly~\eqref{E:gravAnomaly}.
\begin{enumerate}
\item The diffeomorphism variation,
\begin{equation*}
\delta_{\xi}W_2 = - \frac{\tilde{c}_2}{192\pi}\int \partial_{\mu}\xi^{\nu}d\Gamma^{\mu}{}_{\nu}\,,
\end{equation*}
is identical in form to the gravitational anomaly in ordinary $2d$ CFT, with $\tilde{c}_2 = c_L - c_R$ and $c_{L,R}$ the left and right central charges. This anomaly may then be described via anomaly inflow under a suitable extension of $\Gamma^{\mu}{}_{\nu}$ to a higher-dimensional spacetime. This inflow does not require a Lorentzian structure, as is already known from the study of the fractional quantum Hall effect (see e.g.~\cite{Gromov:2015fda}).
\item Concordantly, the diffeomorphism anomaly follows from a descent procedure. The pure boost and mixed boost/Weyl anomalies do not.
\item In ordinary $2d$ CFT, one may redefine the CFT currents in such a way as to shift the gravitational anomaly from a non-diffeomorphism-invariance of the theory to a non-invariance under local Lorentz rotations~\cite{Bardeen:1984pm}. For warped geometry, the frame bundle may be trivialized and no such redefinition exists.
\end{enumerate}

\subsection{Anomalous Ward identities}

Under the assumption that there are no anomalies beyond two derivatives, the anomalous variation of $W$ is given by a sum of the three anomalies we found above,
\begin{align}
\begin{split}
\delta_{\chi}W = &\frac{ \tilde{k}}{8\pi} \int d^2x \sqrt{\gamma}\,\psi + \frac{\tilde{c}_1}{24\pi} \int d^2x \sqrt{\gamma} \,\left( - \dot{\sigma} N + \frac{\psi}{2}N^2\right) 
\\
& \quad -\frac{ \tilde{c}_2}{192\pi} \int \left( \partial_{\mu}\xi^{\nu} d\Gamma^{\mu}{}_{\nu} - d\psi \wedge \Gamma^{\mu}{}_{\nu}w^{\nu}n_{\mu} + d\sigma \wedge \Gamma^{\mu}{}_{\nu}v^{\nu}n_{\mu}\right)\,.
\end{split}
\end{align}
In the parameterization~\eqref{E:deltaChiW} of $\delta_{\chi}W$, we have
\begin{align}
\begin{split}
\mathcal{T}^{\mu}{}_{\nu} &= -\frac{\tilde{c}_2}{96}\varepsilon^{\rho\sigma} \partial_{\rho}\Gamma^{\mu}{}_{\nu\sigma}\,,
\\
\mathcal{P} &= \frac{\tilde{k}}{4} + \frac{\tilde{c}_1}{24}N^2 + \frac{\tilde{c}_2}{96} \varepsilon^{\rho\sigma}\partial_{\rho} \left(\Gamma^{\mu}{}_{\nu\sigma}w^{\nu}n_{\mu}\right) \,,
\\
\mathcal{A} & = \frac{\tilde{c}_1}{12} \left( \dot{N} + N H\right) - \frac{\tilde{c}_2}{96} \varepsilon^{\rho\sigma}\partial_{\rho}\left( \Gamma^{\mu}{}_{\nu\sigma}v^{\nu}n_{\mu}\right)\,,
\end{split}
\end{align}
where $N$ and $H$ were defined in~\eqref{E:defineNH}, $N=\varepsilon^{\mu\nu}\partial_{\mu}n_{\nu}, \,H = \varepsilon^{\mu\nu}\partial_{\mu}h_{\nu}$.

The background fields $(n_{\mu},h_{\nu})$ are conjugate to operators $(\mathcal{N}^{\mu},\mathcal{H}^{\nu})$ which we define through variation via~\eqref{E:deltaW}. The reparameterization, boost, and Weyl symmetries lead to Ward identities.

Plugging~\eqref{E:deltaChiNC} into~\eqref{E:deltaW} and integrating by parts, we find up to a boundary term
\begin{align}
\begin{split}
\delta_{\chi}W & =-\frac{1}{2\pi} \int d^2x \sqrt{\gamma}  \left\{ \sigma n_{\mu}\mathcal{N}^{\mu} - \psi n_{\mu}\mathcal{H}^{\mu}\right.
\\
& \qquad \left.+ \xi^{\mu}\left( \varepsilon_{\mu\nu}(N \mathcal{N}^{\nu}+H \mathcal{H}^{\nu}) - n_{\mu} \frac{1}{\sqrt{\gamma}}\partial_{\nu}(\sqrt{\gamma}\mathcal{N}^{\nu}) - h_{\mu}\frac{1}{\sqrt{\gamma}}\partial_{\nu}(\sqrt{\gamma}\mathcal{H}^{\nu})\right)\right\}\,.
\end{split}
\end{align}
Because the derivative $D_{\mu}$ has torsion, the last two terms above are not covariant divergences. In any case, setting this equal to the anomalous variation~\eqref{E:deltaChiW} we find the anomalous Ward identities
\begin{align}
\begin{split}
\label{E:anomWard}
\frac{1}{\sqrt{\gamma}}\partial_{\mu}(\sqrt{\gamma}\mathcal{N}^{\mu}) - h_{\mu}(N \mathcal{N}^{\mu}+H \mathcal{H}^{\mu}) &= -\frac{1}{\sqrt{\gamma}}v^{\mu}\partial_{\nu}(\sqrt{\gamma}\mathcal{T}^{\nu}{}_{\mu})\,,
\\
 \frac{1}{\sqrt{\gamma}}\partial_{\mu}(\sqrt{\gamma}\mathcal{H}^{\mu}) +n_{\mu}(N\mathcal{N}^{\mu}+H \mathcal{H}^{\mu}) & = -\frac{1}{\sqrt{\gamma}}w^{\mu}\partial_{\nu}(\sqrt{\gamma}\mathcal{T}^{\nu}{}_{\mu})\,,
 \\
 n_{\mu}\mathcal{H}^{\mu} &= \mathcal{P}\,,
 \\
 n_{\mu}\mathcal{N}^{\mu} & =-\mathcal{A}\,.
\end{split}
\end{align}
The boost and Weyl Ward identities fix $n_{\mu}\mathcal{H}^{\mu}$ and $n_{\mu}\mathcal{N}^{\mu}$ in terms of the spacetime background. So there are two free components of $(\mathcal{N}^{\mu},\mathcal{H}^{\mu})$, namely
\beq
\tilde{T} \equiv h_{\mu}\mathcal{N}^{\mu}\,, \qquad \tilde{P} \equiv h_{\mu}\mathcal{H}^{\mu}\,.
\eeq
Soon, we will see that these operators are the $T$ and $P$ of a WCFT.

In flat space with $n=dx^-, \,h=dx^+$, the anomalous Ward identities become
\beq
\partial_+ \tilde{T} = 0\,, \qquad \partial_+ \tilde{P} = 0\,, \qquad \mathcal{H}^- = \frac{ \tilde{k}}{4}\,, \qquad \mathcal{N}^- = 0\,,
\eeq
so that $\tilde{T}=\mathcal{N}^+$ and $\tilde{P}=\mathcal{H}^+$ are right-moving.

\subsection{The relation to central charges}
\label{S:fromAnomToCentral}

In $2d$ CFT, the anomaly coefficients are related to central charges. The Weyl and gravitational anomaly coefficients determine the Virasoro central charges, and the anomalies of any global symmetries determine the levels of the corresponding Kac-Moody algebras. There are many ways to derive these relations. One is to compute the variation of operators under anomalous symmetry transformations, and specialize to the symmetry transformations which take flat space to itself. For the stress tensor, this anomalous transformation law recovers the Schwarzian derivative, and so matches anomalies to central charges.

We adapt this approach to relate the WCFT anomalies to its central charges. The final result is given in~\eqref{E:tildecToc}. Along the way we find the warped analogue of the Polyakov action.

As we mentioned in Subsection~\ref{S:flatWCFT}, the transformation of $T$ and $P$ under the WCFT global symmetries
\begin{equation*}
x^- = x^-(X^-)\,, \qquad x^+ = X^+ + f(X^-)\,,
\end{equation*}
was previously computed in~\cite{Detournay:2012pc}. It is
\begin{align*}
\label{E:newTP}
\hat{T}(X) & = \left( \frac{\partial x^-}{\partial X^-}\right)^2 \left( T(x) - \frac{c}{12}\{  X^-(x^-),x^-\}\right) + \frac{\partial x^-}{\partial X^-}\frac{\partial x^+}{\partial X^-}P(x) - \frac{k}{4}\left( \frac{\partial x^+}{\partial X^-}\right)^2\,,
\\
\hat{P}(X) & = \frac{\partial x^-}{\partial X^-}P(x) - \frac{k}{2}\frac{\partial x^+}{\partial X^-}\,.
\end{align*} 
where 
\begin{equation*}
\{ f(x),x\} = \frac{f'''(x)}{f'(x)} - \frac{3}{2}\left(\frac{f''(x)}{f'(x)}\right)^2\,.
\end{equation*}
is the Schwarzian derivative.

We wish to recover these transformations using the anomalous symmetries. To do so, consider $W$ on a particular, fixed background $(n_{\mu},h_{\nu})$ described with coordinates $x^{\mu}$. Consider  another fixed background $(\hat{n}_{\mu},\hat{h}_{\nu})$, described with coordinates $X^{\mu}$, which can be reached by a coordinate reparamterization, followed by Weyl rescaling $\tau$, followed by a boost $\Psi$, all of which are continuously connected to the identity. In other words,
\beq
x^{\mu} = x^{\mu}(X^{\nu})\,, \qquad \hat{n}_{\mu}(X) = e^{\tau}n_{\nu} \frac{\partial x^{\nu}}{\partial X^{\mu}}\,, \qquad \hat{h}_{\mu}(X) = \left( h_{\nu} - \Psi e^{\tau} n_{\nu}\right)\frac{\partial x^{\nu}}{\partial X^{\mu}}\,,
\eeq
with the transformations non-singular. Then denoting
\beq
\hat{W} = W[\hat{n}_{\mu},\hat{h}_{\mu}; X^{\rho}] \,, \qquad W = W[n_{\mu},h_{\nu};x^{\rho}]\,,
\eeq
we have
\beq
\label{E:W'}
\hat{W} = W + \mathcal{W}\,,
\eeq
where $\mathcal{W}$ accounts for the anomalies. Essentially, the anomalous variation integrates to $\mathcal{W}$. $\mathcal{W}$ is a particular Wess-Zumino term for the anomalies, fixed by the requirement that it vanishes for $x^{\mu}=X^{\mu}, \tau=\Psi = 0$. Alternatively, it is an anomaly effective action. In Appendix~\ref{A:WA} we find 
\begin{align}
\label{E:WA}
\nonumber
\mathcal{W} = &\frac{ \tilde{k}}{8\pi}\int d^2x \sqrt{\gamma} \,\Psi e^{\tau} +\frac{ \tilde{c}_1}{24\pi} \int d^2x \sqrt{\gamma} \left\{ - \dot{\tau} N + \frac{1}{2}\dot{\tau}\tau' + \frac{\Psi e^{\tau}}{2}\left( N - \tau'\right)^2\right\}
\\
& \qquad + \frac{\tilde{c}_2}{192\pi} \Big\{ N_3(g) + \int dg^{\mu}{}_{\nu} (g^{-1})^{\nu}{}_{\rho} \wedge \Gamma^{\rho}{}_{\mu}
\\
\nonumber
 & \qquad \qquad \quad +\int \left\{ \left( v^{\nu} + \Psi e^{\tau}w^{\nu}\right)n_{\mu}d\tau - w^{\nu}n_{\mu}d\left( \Psi e^{\tau}\right) \right\} \wedge \left( \Gamma^{\mu}{}_{\nu} + dg^{\mu}{}_{\rho}(g^{-1})^{\rho}{}_{\nu}\right)\Big\}\,,
\end{align}
where we denote
\beq
g^{\mu}{}_{\nu} = \frac{\partial x^{\mu}}{\partial X^{\nu}}\,, \qquad (g^{-1})^{\mu}{}_{\nu} = \frac{\partial X^{\mu}}{\partial x^{\nu}}\,,
\eeq
and $N_3$ is the usual Wess-Zumino term for a gravitational anomaly. To write it covariantly, we extend the spacetime manifold $\mathcal{M}$ to a three-manifold $\mathcal{N}$ with $\mathcal{M}$ as its boundary, and further extend $g^{\mu}{}_{\nu}$ to a map on $\mathcal{N}$. $N_3$ is a topological term on $\mathcal{N}$,
\beq
N_3(g) = \frac{1}{3}\int_{\mathcal{N}} \text{tr} \left( dg\, g^{-1}\right)^3\,.
\eeq
To proceed, we vary $\hat{W}$ with respect to $(n_{\mu},h_{\nu})$,
\begin{align}
\begin{split}
\delta \hat{W} &=- \frac{1}{2\pi}\int d^2X \sqrt{\gamma'} \left\{ \hat{\mathcal{N}}^{\mu} \delta \hat{n}_{\mu} +\hat{\mathcal{H}}^{\mu} \delta \hat{h}_{\mu} \right\} 
\\
&=-\frac{1}{2\pi} \int d^2x \sqrt{\gamma} e^{\tau} \left\{ e^{\tau} \delta n_{\mu}\,  g^{\mu}{}_{\nu}\hat{\mathcal{N}}^{\nu} + \left( \delta h_{\mu} - \Psi e^{\tau} \delta n_{\mu}\right) g^{\mu}{}_{\nu}\hat{\mathcal{H}}^{\nu}\right\}\,.
\end{split}
\end{align}
By~\eqref{E:W'} this is equal to the variation of $W+\mathcal{W}$. Decomposing
\beq
\delta \mathcal{W} = -\frac{1}{2\pi}\int d^2x \sqrt{\gamma} \left\{  \mathcal{N}_A^{\mu} \delta n_{\mu}+  \mathcal{H}_A^{\mu}\delta h_{\mu}\right\}\,,
\eeq
which gives the transformation laws of the currents $(\mathcal{N}^{\mu},\mathcal{H}^{\nu})$ to be
\begin{align}
\begin{split}
g^{\mu}{}_{\nu}\hat{\mathcal{N}}^{\nu}(X) & = e^{-2\tau}\left( \mathcal{N}^{\mu}(x) +\mathcal{N}_A^{\mu}\right) +e^{-\tau} \Psi \left(\mathcal{H}^{\mu}(x) + \mathcal{H}_A^{\mu}\right)\,,
\\
g^{\mu}{}_{\nu}\hat{\mathcal{H}}^{\nu}(X) & = e^{-\tau}\left(  \mathcal{H}^{\mu}(x) + \mathcal{H}^{\mu}_A\right) \,.
\end{split}
\end{align}
Contracting these vectors with $h_{\mu}-\Psi e^{\tau}n_{\mu}$ and using the anomalous Ward identities~\eqref{E:anomWard} we have
\begin{align}
\nonumber
\hat{\tilde{T}}(X) & = e^{-2\tau} \left( \tilde{T}(x) +h_{\mu}\mathcal{N}_A^{\mu}\right) + e^{-\tau} \Psi \left( \tilde{P}(x) + \mathcal{A} - n_{\mu}\mathcal{N}_A^{\mu}+ h_{\mu}\mathcal{H}_A^{\mu}\right)-\Psi^2 \left( \mathcal{P} + n_{\mu}\mathcal{H}_A^{\mu}\right)\,,
\\
\label{E:TPnew}
\hat{\tilde{P}}(X) & = e^{-\tau} \left(\tilde{P}(x) + h_{\mu}\mathcal{H}_A^{\mu}\right)-\Psi \left(  \mathcal{P}+n_{\mu}\mathcal{H}_A^{\mu}\right)\,.
\end{align}

We compute the currents $(\mathcal{N}_A^{\mu},\mathcal{H}_A^{\nu})$ in Appendix~\ref{A:WA}. For the transformations which take the flat structure $n=dx^-, h=dx^+$ to itself, which we remind the reader are given in~\eqref{E:WCFTsymmetries2},
\begin{equation*}
x^- = x^-(X^-)\,, \qquad x^+ = X^+ + f(X^-)\,, \qquad e^{\tau} = \left( \frac{\partial x^-}{\partial X^-}\right)^{-1}\,, \qquad \Psi = \frac{\partial x^+}{\partial X^-}\,,
\end{equation*}
these currents simplify to
\begin{align}
\begin{split}
\mathcal{N}_A^{\mu} & =- \frac{\tilde{k}}{4} \Psi e^{\tau} v^{\mu} +\frac{\tilde{c}_1}{12} \left( \varepsilon^{\mu\nu}\partial_{\nu}\dot{\tau} + \frac{\dot{\tau}^2}{2}w^{\mu}\right)\,,
\\
\mathcal{H}_A^{\mu} & = - \frac{\tilde{k}}{4} \Psi  e^{\tau}w^{\mu} \,.
\end{split}
\end{align}
In particular,
\beq
h_{\mu}\mathcal{N}_A^{\mu} = -\frac{\tilde{c}_1}{12} \left( \ddot{\tau} - \frac{\dot{\tau}^2}{2}\right)= -\frac{\tilde{c}_1}{12} \{ X^-(x^-),x^-\}\,.
\eeq
Using that $\mathcal{P}=\frac{\tilde{k}}{4}$ and $\mathcal{A}=0$ in flat space,~\eqref{E:TPnew} becomes
\begin{align}
\begin{split}
\label{E:finalWtransformation}
\hat{\tilde{T}}(X) & = \left( \frac{\partial x^-}{\partial X^-}\right)^2 \left( \tilde{T}(x)  - \frac{\tilde{c}_1}{12} \{ X^-(x^-),x^-\}\right) + \frac{\partial x^-}{\partial X^-}\frac{\partial x^+}{\partial X^-}\tilde{P}(x)- \frac{\tilde{k}}{4} \left( \frac{\partial x^+}{\partial X^-}\right)^2\,,
\\
\hat{\tilde{P}}(X) & = \left( \frac{\partial x^-}{\partial X^-}\right) \tilde{P}(x)  -\frac{ \tilde{k}}{2} \frac{\partial x^+}{\partial X^-}\,.
\end{split}
\end{align}
This matches the transformations~\eqref{E:newTP} of $T$ and $P$ once we identify
\beq
\tilde{T} = T\,, \qquad \tilde{P} = P\,,
\eeq
and match the central charges as
\beq
\label{E:tildecToc}
\tilde{k} = k\,, \qquad \tilde{c}_1 = c\,.
\eeq
This is the main result of this Subsection.

Note that the gravitational anomaly coefficient $\tilde{c}_2$ completely drops out of the warped transformation law~\eqref{E:finalWtransformation}.

\section{Thermodynamics}
\label{S:thermodynamics}

We continue with a discussion of the thermodynamics of local WCFTs. Much of it involves the hydrostatic partition function~\cite{Jensen:2012jh,Banerjee:2012iz} (see also refs.~\cite{Jensen:2014ama,Jensen:2014wha} which adapt the hydrostatic machinery to Galilean theories).

Our main result in this Section is to use the warped anomalies obtained in Section~\ref{S:anomalies} to derive the warped thermodynamics. Because we use the anomalies, our analysis holds for local WCFTs. We thereby derive the warped ``Cardy formula'' for local WCFTs, recovering the result of Detournay, Hartman, and Hofman previously obtained in~\cite{Detournay:2012pc}.

Before proceeding, should we regard $x^-$ or $x^+$ as the ``time direction?'' Meaning, is the Hamiltonian that appears in the Boltzmann weight $\exp(-\beta H)$ the generator of translations in $x^-$ or in $x^+$? On the one hand, the flat space Ward identities
\begin{equation*}
\partial_+ T = \partial_+ P = 0\,,
\end{equation*}
suggest that $x^+$ should be regarded as ``time.'' Relatedly, the Virasoro and Kac-Moody modes $L_n$ and $J_n$ may be obtained by from the modes of $T$ and $P$ around a compact $x^-$ circle. On the other hand, in Galilean theories, it is natural to identify time from $n = dx^-$. Further, correlation functions have polynomial falloff in the $x^-$ direction, as they do in conformal quantum mechanics.

We proceed by being agnostic and covering both possibilities. Consider a thermal state characterized by an (unnormalized) density matrix
\beq
\rho = e^{- \beta \big(a^- P_- + a^+ P_+\big)}\,,
\eeq
where $P_{\pm}$ generates translations in $x^{\pm}$. When $a^-$ is nonzero, we may always perform a Galilean boost to eliminate $a^+$:
\beq
\label{E:rho1}
\rho \to \rho' = e^{- \beta a^- P_-'}\,.
\eeq
However, when $a^-$ vanishes,
\beq
\label{E:rho2}
\rho = e^{-\beta a^+ P_+}\,,
\eeq
is boost-invariant. The density matrices~\eqref{E:rho1} and~\eqref{E:rho2} correspond respectively to $x^-$ and $x^+$ being the ``time direction.''

The latter is problematic. When the $x^-$ circle is compact with length $2\pi$, $P_+$ is the zero mode of the Kac-Moody algebra,
\beq
P_+ = \frac{1}{2\pi}\int_0^{2\pi} dx^- P(x) = J_0\,,
\eeq
whose spectrum is generally not bounded below.

Given a theory with a functional integral description, the thermal partition function is related to a suitable Euclidean partition function. The Boltzmann weight $\exp(-\beta H)$ generates a translation $\beta$ in imaginary time. Its trace corresponds to the functional integral on a Euclideanized spacetime where imaginary time is compacitified with periodicity $\beta$. For a theory with a gravity dual, the imaginary time circle is contractible in the dual (Euclidean) black hole geometry, shrinking to zero size at the Euclideanized horizon.

In their derivation of a warped analogue of the Cardy formula, Detournay, Hartman, and Hofman~\cite{Detournay:2012pc} studied thermal states of the form~\eqref{E:rho1} rather than~\eqref{E:rho2}. They also considered asymptotically warped AdS$_3$ black holes putatively dual to WCFTs at nonzero temperature, for which the $x^-$ circle is contractible in the dual geometry.

Given the problem with the mixed state~\eqref{E:rho2} and in order to make contact with Detournay, Hartman, and Hofman, we treat $x^-$ as the time direction and consider thermal states of the form $\rho = \exp(- \beta P_-)$.

More generally, we consider hydrostatic states. These are spatially varying, but time-independent equilibria supported by a time-independent, but spatially varying background spacetime. The equilibrium state is hydrostatic when the spatial variations are much longer than the inverse temperature. Suppose that the correlation length of the WCFT is finite (later we will see that this assumption is correct). Then the Euclidean partition function in the time independent background -- the hydrostatic partition function -- can be written locally on a constant time slice in a gradient expansion. See~\cite{Jensen:2012jh,Banerjee:2012iz,Jensen:2014ama,Jensen:2014wha} for details. 

\subsection{Hydrostatic equilibria}

Hydrostatic states are characterized by time-independent geometries. More covariantly, there is a timelike vector field $K^{\mu}$ (timelike means $K^{\mu}n_{\mu}>0$ in this context) and boost $\psi_K$ which generates a symmetry of the background: 
\begin{align}
\begin{split}
\delta_K n_{\mu} &= \pounds_K n_{\mu} = 0\,,
\\
\delta_K h_{\mu} & = \pounds_K h_{\mu} - \psi_K n_{\mu}= 0\,.
\end{split}
\end{align}
We can always pick a choice of coordinates $(t,x)$ such that $K^{\mu}\partial_{\mu} = \partial_t$ and a boost ``gauge'' so that $\psi_K=0$. In this ``static gauge'' the background is explicitly time-independent,
\begin{align}
\begin{split}
\label{E:staticGauge}
n_{\mu} dx^{\mu} & =  n_0(x) (dt + \mathfrak{a}(x)dx)\,,
\\
h_{\mu}dx^{\mu} & = u(x)n_0(x) (dt + \mathfrak{a}(x)dx) + h_1(x)dx\,,
\end{split}
\end{align}
and we take $n_0,h_1>0$. The Euclideanized theory lives on this background after Wick-rotating $t = i t_E$ and identifying $t_E \sim t_E + \beta$. The Euclidean partition function $\mathcal{Z}_E=\mathcal{Z}_E [n_0,a,u,h_1] $ may be understood as an ordinary thermal partition function
\beq
\mathcal{Z}_E = \text{tr}\Big( e^{- \beta \mathcal{H}}\Big)\,,
\eeq
where $\mathcal{H}$ generates translations in $t$. The hydrostatic generating functional is
\beq
W_{hydrostatic} = - \ln \mathcal{Z}_E\,,
\eeq
and the currents of the hydrostatic state are obtained by functional variation.

The static gauge does not completely fix the coordinate/boost/Weyl gauge. The residual freedom is the combination of the time-independent coordinate transformation
\beq
x = x(X)\,, \qquad t = T + f(X)\,,
\eeq
along with time-independent boost/Weyl rescalings. Under spatial reparameterizations $x = x(X)$, $( n_0, \mathfrak{a},u)$ are invariant while $h_1$ transforms like a measure. Under additive reparameterizations of time (a ``thermal Kaluza-Klein'' $U(1)$ transformation) $t = T + f(x)$, $(n_0,u,h_1)$ are invariant while $\mathfrak{a}$ transforms like a connection,
\beq
\mathfrak{a}(x) \to \mathfrak{a}(x) +\partial_x f(x),.
\eeq
Under boosts and Weyl rescalings, we have
\begin{align}
\begin{split}
\text{Boost:} \quad & u \to u - \psi \,,
\\
\text{Weyl:} \quad& n_0 \to e^{\sigma}n_0\,, \qquad u\to e^{-\sigma} u\,,
\end{split}
\end{align}
with all other fields neutral.

Most of this may be understood in a manifestly covariant way. Defining a local temperature $T_L$ (as opposed to $T$, which we are using to notate the stress tensor) and velocity $u^{\mu}$ (satisfying $u^{\mu}n_{\mu}=1$),
\beq
T_L = \frac{1}{\beta n_{\mu}K^{\mu}}\,, \qquad u^{\mu}  = \frac{K^{\mu}}{n_{\nu}K^{\nu}}\,,
\eeq
we see that $n_0$ and $u$ are spacetime scalars,
\beq
n_0 = \frac{1}{\beta T_L}\,, \qquad u = u^{\mu}h_{\mu}\,.
\eeq
Meanwhile, under the convention that $n_0,h_1>0$,
\beq
\sqrt{\gamma} = n_0 h_1\,,
\eeq
so $h_1$ gives a good measure for spatial integrals.

\subsection{The hydrostatic partition function and equilibrium currents}

$W_{hydrostatic}$ is strongly constrained by the symmetries of the problem. First, it may be written locally in $x$ in a gradient expansion,
\beq
W_{hydrostatic} = \beta \int dx \sqrt{\gamma}\left( \mathcal{L}_0 + \mathcal{L}_1 +\hdots \right)\,,
\eeq
where (up to anomalies) $\mathcal{L}_n$ are boost/KK-invariant scalars containing $n$ derivatives which scale with weight $-1$ under Weyl rescalings. Second, under time-independent symmetry transformations, it ought to reproduce the anomalies of the microscopic theory. Parameterizing the anomalies as in~\eqref{E:deltaChiW}, we have
\beq
\delta_{\chi}W_{hydrostatic} = \frac{\beta}{2\pi} \int  dx \sqrt{\gamma} \left\{ \partial_{\mu} \xi^{\nu} \mathcal{T}^{\mu}{}_{\nu} + \psi \mathcal{P} + \sigma \mathcal{A}\right\}\,,
\eeq
where the anomalies $(\mathcal{T}^{\mu}{}_{\nu},\mathcal{P},\mathcal{A})$ are evaluated in the hydrostatic background~\eqref{E:staticGauge}.

Following previous work (e.g.~\cite{Jensen:2013kka,Jensen:2013rga}), we decompose the hydrostatic generating functional into a sum of three therms:
\beq
W_{hydrostatic} = W_0 + W_A + W_{trans}\,.
\eeq
The microscopic anomalies are reproduced by the functional $W_A$, $W_{trans}$ refers to the set of Chern-Simons terms on a constant time slice, and $W_0$ to everything else. In relativistic field theory, $W_{trans}$ is fixed by the underlying anomalies and its variations encode ``transcendental anomaly-induced transport.'' We find a similar story for WCFT shortly.

Let us parameterize the most general $W_{hydrostatic}$ to zeroth order in derivatives. The ``anomaly action'' $W_A$ has three parts, one for each anomaly. It is
\beq
W_A = -\frac{k\beta}{8\pi} \int dx \sqrt{\gamma}\, u + \mathcal{O}(\partial^2)\,.
\eeq
The unique Chern-Simons term on the constant-time slice is
\beq
W_{trans} = \frac{\tilde{C}}{2\pi\beta}\int dx \,\mathfrak{a}\,,
\eeq
where $\tilde{C}$ is a real constant and the factor of $\beta$ follows from dimensional analysis. Finally,
\beq
W_0  = \frac{\beta}{2\pi} \int dx \sqrt{\gamma}\, (p T_L) + \mathcal{O}(\partial)\,,
\eeq
where $p$ is a real constant.

The real-time currents of the hydrostatic state are
\begin{align}
\begin{split}
\label{E:hydrostaticCurrents}
T(x) & =- \tilde{C} T_L^2 + p T_L u - \frac{k}{4}u^2 + \mathcal{O}(\partial)\,,
\\
P(x) & =- p T_L + \frac{k}{2}u + \mathcal{O}(\partial)\,.
\end{split}
\end{align}
In flat space with constant $T_L, u$, we extract the expectation values of the zero-modes $L_0$ and $J_0$ of the Virasoro and Kac-Moody algebras by integrating $-T$ and $P$ over the thermal circle,
\begin{align}
\begin{split}
\label{E:L0J0}
L_0 -\frac{c}{24}&=\tilde{C} T_L^2 - p T_L u + \frac{k}{4}u^2  \,, 
\\
J_0 & =\frac{k}{2}u - p T_L\,.
\end{split}
\end{align}
 We also obtain the hydrostatic entropy current $s^{\mu}$\footnote{This entropy current, stress tensor, and momentum current saturate the local Second law
\beq
T_L\frac{1}{\sqrt{\gamma}}\partial_{\mu} (\sqrt{\gamma}s^{\mu}) + \mathcal{D}_T + u \mathcal{D}_P \geq 0\,,
\eeq
where $\mathcal{D}_T$ and $\mathcal{D}_P$ are the Ward identities~\eqref{E:anomWard} involving the divergence of $\mathcal{N}^{\mu} = - \mathcal{A}v^{\mu}+ T(x) w^{\mu} $ and $\mathcal{H}^{\mu} = \mathcal{P}v^{\mu} + P(x)w^{\mu} $ respectively.}
\beq
s^{\mu} =- \frac{p}{2\pi} u^{\mu} + \frac{\tilde{C}}{\pi}T_L w^{\mu}\,.
\eeq

\subsection{ From the plane to the cylinder and the warped Cardy formula}

The constants $(p,\tilde{C})$ are a priori unrelated to the anomaly coefficients. However, as in $2d$ CFT, there is a conformal transformation from the plane (which here is $n=dx^-, h=dx^+$) to the thermal cylinder. It is
\beq
\label{E:planeToCylinder}
x^- = \frac{\beta}{2\pi}\exp\left( \frac{2\pi i t_E}{\beta}\right)\,, \qquad x^+ = x - i u t_E\,,
\eeq
where $t_E \sim t_E + \beta$. The resulting thermal state has a temperature $T_L = 1/\beta$ and carries a velocity $u$.\footnote{The velocity $u$ corresponds to $-2\alpha$ in~\cite{Detournay:2012pc}.} One expects the currents $T$ and $P$ to generally vanish on the plane, but for certain theories it could be that they do not, in which case we can parameterize them as
\beq
\langle T(x^-)\rangle = \frac{\ell}{(x^-)^2}\,, \qquad \langle P(x^-)\rangle =- \frac{i q}{x^-}\,.
\eeq
(We have in mind that $q$ is pure imaginary. Bear with this unusual convention for a moment.) When then obtain $T$ and $P$ on the thermal cylinder from the warped transformation law~\eqref{E:finalWtransformation}. Analytically continuing back to real time we have
\begin{align}
\begin{split}
\hat{T}(x) & =4\pi^2\left( \frac{c}{24}-\ell\right) T_L^2 +2\pi i q T_L u- \frac{k}{4}u^2\,,
\\
\hat{P}(x) & = \frac{k}{2}u- 2\pi i qT_L \,.
\end{split}
\end{align}
Matching to~\eqref{E:hydrostaticCurrents} we see that $(p,\tilde{C})$ are fixed as
\beq
\label{E:pC}
p = 2\pi i q\,, \qquad \tilde{C} = 4\pi^2\left(\frac{c}{24}-\ell\right)\,.
\eeq

In fact, any hydrostatic background (in infinite volume) can be reached from the combination of the map~\eqref{E:planeToCylinder} from the plane to the cylinder, followed by a time-independent Weyl transformation and boost. The hydrostatic partition function and hydrostatic currents can thereby be obtained by integrating the phase picked up by the partition function under this sequence of maps. 

Fixing $p$ and $\tilde{C}$ as in~\eqref{E:pC}, the WCFT hydrostatic partition function is
\beq
W_{hydrostatic} = \frac{2\pi }{\beta}\left( \frac{c}{24}-\ell\right)\int dx\, \mathfrak{a} +iq\beta \int dx \sqrt{\gamma} \,T_L+ \frac{k\beta}{8\pi}\int dx \sqrt{\gamma}\, u + \mathcal{O}(\partial)\,,
\eeq
and the entropy current in the thermal state is
\beq
s^{\mu} =-iqu^{\mu} + 4\pi\left( \frac{c}{24}-\ell\right)T_L w^{\mu} + \mathcal{O}(\partial)\,.
\eeq
These results hold in infinite volume, or in finite volume and high-temperature. Normalizing the spatial volume to be $2\pi$, the total entropy \emph{flux} in flat space is $S =2\pi s^{\mu}h_{\mu} = -2\pi iq u+ 8\pi^2\left( \frac{c}{24}-\ell\right)T_L$, which after using~\eqref{E:L0J0} can be expressed in the microcanonical ensemble as
\beq
\label{E:cardy}
S = 4\pi \sqrt{\left( \frac{c}{24}-\ell + \frac{q^2}{k}\right)\left( L_0 - \frac{J_0^2}{k}-\frac{c}{24}\right)} - \frac{4\pi iJ_0 q}{k}\,.
\eeq
Observe that for a WCFT which is translationally invariant in the plane, both $\ell$ and $q$ vanish, in which case this becomes the standard Cardy formula for a chiral CFT with a $U(1)$ global symmetry,
\beq
S \to 4\pi \sqrt{\frac{c}{24}\left( L_0 - \frac{J_0^2}{k}-\frac{c}{24}\right)}\,.
\eeq

This result is precisely the warped Cardy formula obtained by Detournay, Hartman, and Hofman~\cite{Detournay:2012pc}, 
\beq
S_{DH^2} = 4\pi \sqrt{-\left( L_0^{vac}-\frac{(J_0^{vac})^2}{k}\right) \left( L_0 - \frac{J_0^2}{k}-\frac{c}{24}\right)} - \frac{4\pi i J_0 J_0^{vac}}{k}\,,
\eeq
where $L_0^{vac}$ and $J_0^{vac}$ are the expectation values of $L_0$ and $J_0$ in the vacuum on a cylinder, with the identification
\beq
L_0^{vac} = \ell - \frac{c}{24}\,, \qquad J_0^{vac} = q\,.
\eeq
They further argue that
\beq
\ell = \frac{q^2}{k}\,,
\eeq
so that the asymptotic entropy is
\beq
\label{E:DH2cardy}
S_{DH^2} \to 2\pi \sqrt{\frac{c}{6}\left( L_0 - \frac{J_0^2}{k}-\frac{c}{24}\right)} -\frac{4\pi i q J_0}{k}\,.
\eeq
Note that, in terms of an entropy current, this is really the entropy flux rather than the entropy.

We conclude this Section with two minor comments. In writing a local hydrostatic partition function we assumed that the correlation length was finite. Observe that the connected one-point functions of the stress tensor and momentum in a general hydrostatic state may be obtained by the combination of the warped map to the cylinder, followed by a time-independent Weyl rescaling and boost. These one-point functions are the functional variations of the Wess-Zumino term $\mathcal{W}$ that reared its head in Subsection~\ref{S:fromAnomToCentral}; they are completely local. So, as far as the stress tensor and momentum are concerned, the correlation length of a WCFT is not merely finite, it vanishes.

Second, to obtain the constants $p$ and $\tilde{C}$ we used that the plane is conformal to the cylinder. If the WCFT is translationally invariant in the plane, then $\ell = q = 0$ and the coefficients $p$ and $\tilde{C}$ are fixed by the underlying anomalies. A similar computation in $2d$ CFT fixes the pressure and the analogue of $\tilde{C}$ in terms of the total central charge $c_L+c_R$ and gravitational anomaly $c_L - c_R$ respectively (see e.g.~\cite{Jensen:2012kj}). Recently, the authors of~\cite{Golkar:2015oxw} (see also~\cite{Chowdhury:2016cmh,Glorioso:2017lcn}) have pointed out that (the fractional part of) the $2d$ CFT version of $\tilde{C}$ is also fixed by a global gravitational anomaly, and we expect a similar phenomenon here.

The above results for the hydrostatic partition function, hydrostatic currents, and entropy have all been obtained on symmetry grounds alone and so hold for any local WCFT, including the free-field theories presented in Section~\ref{S:freeWCFT}.

\section{Holography}
\label{S:holography}

In this Section we consider WCFTs dual to gravity on so-called warped AdS$_3$ spacetimes. We find three main results. First, the boundary of these spacetimes is not equipped with warped geometry, but instead a variant with $dn=0$. Second, these holographic WCFTs do not possess the warped anomalies we found in Section~\ref{S:anomalies}. Finally, these WCFTs are non-local, and in fact are semi-local in the same sense as in~\cite{Iqbal:2011in}

\subsection{General WAdS$_3$ spacetimes}

We consider spacelike warped AdS$_3$ spacetimes, where a spatial line or circle is fibered over AdS$_2$. There are also timelike and null warped AdS$_3$ spacetimes that we do not consider, as gravity on those spacetimes does not appear to be dual to a WCFT.

The metric on the analogue of the Poincare patch for WAdS$_3$ is
\beq
\label{E:WAdS3}
G_{WAdS_3} = L^2 \left( \frac{dr^2}{r^2} - r^2 (dx^-)^2 + \alpha^2 ( dx^+ + r\, dx^-)^2\right)\,,
\eeq
where $L$ is a radius and $\alpha$ the warping parameter. For $\alpha^2 = 1$ this spacetime is the Poincar\'e patch of AdS$_3$ with a radius of curvature $L^2/4$, written as an $\mathbb{R}$ bundle over AdS$_2$. For $\alpha^2>1$ the fiber is stretched, and for $\alpha^2 <1$ it is squashed. There are spacelike warped black holes for $\alpha^2>1$. In the next Subsection we will see that the dual to gravity on WAdS$_3$ is semi-local, and further that there are dynamical instabilities whenever $\alpha^2>1$.

Spacelike WAdS$_3$ appears as a solution of IIB string theory via a TsT transformation and dimensional reduction of its AdS$_3\times \mathbb{S}^3\times \mathcal{M}_4$ vacuum~\cite{Detournay:2012pc}. It is also a solution to topologically massive gravity in three dimensions with
\beq
S_{TMG} = \frac{1}{16\pi G}\left\{ \int d^3x \sqrt{-g} \left( R +2\right) - \frac{1}{6\nu}\int d^3x \sqrt{-g} \,\varepsilon^{\mu\nu\rho}\left( \Gamma^{\alpha}{}_{\beta\mu}\partial_{\nu}\Gamma^{\beta}{}_{\alpha\rho} + \frac{2}{3}\Gamma^{\alpha}{}_{\beta\mu}\Gamma^{\beta}{}_{\gamma\nu}\Gamma^{\gamma}{}_{\alpha\rho}\right)\right\}\,,
\eeq
and~\cite{Moussa:2003fc,Anninos:2008fx}
\beq
L^2 = \frac{1}{\nu^2+3}\,, \qquad \alpha^2 = \frac{4\nu^2}{\nu^2+3}\,.
\eeq
Perhaps most interestingly, global spacelike WAdS$_3$ arises as a submanifold inside the near-horizon geometry of the extremal Kerr black hole~\cite{Bardeen:1999px,Bengtsson:2005zj} at fixed polar angle.

The ``Poincare patch'' of spacelike WAdS$_3$ has $PSL(2;\mathbb{R})\times \mathbb{R}$ isometry, which is the global subgroup of the warped conformal symmetry. Uncovering the Virasoro and Kac-Moody symmetries requires more work; one must define boundary conditions on the asymptotic form of the metric so that the asymptotic symmetry algebra is that of a Virasoro algebra and abelian Kac-Moody algebra. This was done in~\cite{Compere:2008cv}.

In the usual AdS$_3$/CFT$_2$ dictionary, the CFT dual to gravity on AdS$_3$ is coupled to an external metric $g_{\mu\nu}$ in the following way. Given an asymptotically locally AdS$_3$ spacetime, the conformal boundary is endowed with a conformal structure, which is the equivalence class of $g$ under Weyl rescaling. For WAdS$_3$ spacetimes we expect a similar story where the bulk metric induces a ``warped conformal structure'' on the boundary, a representative of which is given by two independent one-forms $(n_{\mu},h_{\nu})$.

The natural way to accomplish this is to pass to a first-order formalism in the bulk as has been done for Lifshitz holography in e.g.~\cite{Christensen:2013rfa} (and substantially generalized in~\cite{Hartong:2015wxa}). Rewrite the WAdS$_3$ metric in terms of a coframe $e^A=e^A_{M}dx^M$ with $x^M=(x^{\mu},r)$ as
\beq
G_{WAdS_3} = \eta_{AB} e^A e^B\,,
\eeq
with
\beq
\label{E:WAdS3vielbein}
e^0 = r L dx^-\,, \qquad e^1 = \alpha L(dx^+ + r\,dx^-)\,, \qquad e^2 = \frac{L}{r}dr\,.
\eeq
In this form, it appears that the bulk coframe induces two one-forms on the boundary which here would be $dx^-$ and $dx^+$, and it would be tempting to identify these as the $n$ and $h$ of a warped geometry. However this does not quite work.

To see this, recall a minor result we obtained at the end of Subsection~\ref{S:galilean}: every warped geometry is equivalent to the flat one $n=dx^-,\, h=dx^+$ by the combination of a coordinate transformation, Weyl rescaling, and boost. If a bulk coframe on WAdS$_3$ induces a warped conformal structure on its boundary, then there must be a way to use a bulk diffeomorphism and local Lorentz rotation to induce the most general $n$ and $h$ on its boundary, and further to identify the gravity dual of a Weyl rescaling and boost.

Fortunately, we already know how to get the gravity dual of the boundary symmetry transformations, whether they be diffeomorphisms, boosts, or otherwise. After fixing a radial gauge (the analogue of the choice of Fefferman-Graham coordinates in an asymptotically AdS spacetime) and the large$-r$ falloffs of the coframe, one solves for the most general asymptotic form of a bulk diffeomorphism and local Lorentz rotation that is consistent with the gauge choice and falloffs. These are the so-called Penrose-Brown-Henneaux (PBH) transformations~\cite{Imbimbo:1999bj}.

For the case at hand we pick a gauge so that $e^2 =L dr/r$ always, and that $e^0$ and $e^1$ have no components along $dr$. Further, we require that $e^0$ and $e^1$ grow as $r$ in a parallel way, i.e. that they satisfy the boundary condition that
\beq
\label{E:warpedBC}
e^0 =  r L \Big( \mathfrak{n}_{\mu}(x^{\nu}) + \hdots \Big) dx^{\mu} \,, \qquad e^1 = \alpha L r \Big( \mathfrak{n}_{\mu}(x^{\nu}) + \hdots \Big) dx^{\nu}\,,
\eeq
where the dots indicate terms that vanish as $r$ tends to infinity. In this way the asymptotic growth of the coframe only encodes \emph{one} one-form on the boundary. (The other boundary condition where $e^0$ and $e^1$ grow as $r$ in different directions leads to an asymptotically AdS$_3$ spacetime instead.)

In this context, a PBH transformation is a bulk diffeomorphism and Lorentz rotation consistent with this gauge choice and boundary condition. It is easy to show that at the infinitesimal level, under which the bulk coframe varies as
\beq
\delta e^A = \pounds_{\Xi} e^A - v^A{}_B e^B\,,
\eeq
with $\Xi^M$ a bulk diffeomorphism and $v^A{}_B$ a bulk Lorentz rotation, a PBH transformation of pure WAdS$_3$~\eqref{E:WAdS3vielbein} is parameterized by a vector field $\xi^{\mu}(x^{\nu})$, a Weyl rescaling $\sigma(x^{\mu})$, and boost $\psi(x^{\mu})$, and is given by\footnote{We are grateful to J.~Hartong for pointing out an error in a previous version of this computation.}
\begin{align}
\begin{split}
\Xi^- & = \xi^- + \frac{\partial_+ \sigma}{r}-\frac{\partial_- \sigma}{2r^2}\,,
\\
\Xi^+ & = \xi^+ + \frac{\alpha^2-1}{\alpha^2} (\ln r) \partial_+ \sigma  + \frac{\partial_- \sigma}{r}\,,
\\
\Xi^r &= r\, \sigma\,,
\\
v^0{}_1 & = -\frac{\alpha\,\psi}{r}\,,
\\
v^0{}_2 & = \partial_+\sigma-\frac{\partial_-\sigma}{r}\,,
\\
v^1{}_2 & = - \frac{\partial_+\sigma}{\alpha}\,.
\end{split}
\end{align}
Under it the bulk coframe varies as
\begin{align}
\label{E:deltaE}
\delta e^0& =   L \left[ r(d\xi^- + \sigma dx^-) +\alpha^2\psi \left(dx^- + \frac{dx^+}{r}\right) + d(\partial_+\sigma) -\frac{d(\partial_-\sigma)}{2r} \right] \, , 
\\
\nonumber
\delta e^1& = \alpha L \left[ r \Big(  d\xi^- + \sigma dx^- \Big) +\Big( d\xi^+ - \psi dx^-\Big)  +\left( \frac{\alpha^2-1}{\alpha^2}\ln r +1\right)d(\partial_+\sigma)+\frac{d(\partial_-\sigma)}{2r}\right]\,.
\end{align}

Observe that if we fix $e^0$ to $\mathcal{O}(r)$ and $e^1$ to $\mathcal{O}(r^0)$ as boundary conditions, the PBH transformations consistent with those boundary conditions are 
\beq
\xi^- = \xi^-(x^-)\,, \qquad \xi^+ = f(x^-)\,, \qquad \sigma = - \partial_- \xi^-\,, \qquad \psi = \partial_- f\,,
\eeq
which exponentiate to (in the $r\to\infty$ limit)
\beq
x^- = x^-(X^-)\,, \qquad x^+ = X^+ + f(X^-)\,.
\eeq
Subject to these boundary conditions, these diffeomorphisms generate the asymptotic symmetry group of WAdS$_3$. We recognize them as the warped conformal symmetries of a WCFT as we discussed in Section~\ref{S:flatWCFT}.

It is tempting to identify a boundary $n_{\mu}$ and $h_{\nu}$ from the $\mathcal{O}(r)$ part of $e^0$ and the $\mathcal{O}(r^0)$ part of $e^1$ respectively. (More precisely, to identify $h_{\mu}$ from the $r\to\infty$ limit of $e^0-\alpha e^1$.) The $\mathcal{O}(r)$ part of $\delta e^0$ is exactly how the one-form $n_{\mu}$ of a warped geometry varies under a diffeomorphism $\xi^{\mu}$, Weyl rescaling $\sigma$, and boost $\psi$ when it is initially $n = dx^-$. The second set of parenthesis in $\delta e^1$ in~\eqref{E:deltaE} is exactly how $h_{\mu}$ varies under the same transformation when it is initially $dx^+$. However, the logarithm in $\delta e^1$ spoils this identification (note that there is no logarithm when $\alpha^2=1$: however in this case the geometry becomes asymptotically AdS$_3$ rather than WAdS$_3$).

In fact the situation is worse. Exponentiating the PBH transformation, the logarithmic term exponentiates to a power of $r$ that depends on $\partial_+\sigma$.

In order to identify two boundary one-forms from the bulk coframe, we then require the additional condition that the PBH transformations satisfy $\partial_+\sigma = 0$ so that there is no logarithmic term in $e^1$. Comparing the variation $\delta e^0$ and the identification of the boundary $\mathfrak{n}_{\mu}$ from the asymptotic growth of the coframe as in~\eqref{E:warpedBC}, this extra condition amounts to
\beq
d\mathfrak{n} = 0\,,
\eeq 
and further that Weyl rescalings $\sigma$ retain $d\mathfrak{n}=0$.

We can then identify two boundary one-forms $(\mathfrak{n}_{\mu},\mathfrak{h}_{\nu})$ via
\beq
e^0 =  r L \left( \mathfrak{n}_{\mu}(x^{\nu})dx^{\mu} + \hdots\right)\,, \qquad e^1 = L \alpha \left( r \,\mathfrak{n}_{\mu}(x^{\nu})dx^{\mu} + \mathfrak{h}_{\mu}(x^{\nu}) dx^{\mu} + \hdots \right)\,.
\eeq
All in all we see that a WAdS$_3$ geometry does not induce a warped geometry on its boundary, but something else. It induces a Newton-Cartan structure $(\mathfrak{n}_{\mu},\mathfrak{h}_{\nu})$ subject to $d\mathfrak{n}=0$, along with a Weyl rescaling symmetry $\Omega$ (satisfying $\mathfrak{w}^{\mu}\partial_{\mu}\Omega = 0$, where $\mathfrak{w}^{\mu}n_{\mu}=0$ and $\mathfrak{w}^{\mu}\mathfrak{h}_{\mu}=1$) and a boost symmetry $\psi$ under which
\beq
\mathfrak{n} \to e^{\Omega} \mathfrak{n}\,, \qquad \mathfrak{h} \to \mathfrak{h} - \psi\,\mathfrak{n}\,.
\eeq
This structure is identical to warped geometry, up to the condition that $d\mathfrak{n}=0$.

\subsection{Anomaly non-matching}

In Section~\ref{S:anomalies} we found that WCFTs possess a boost anomaly and a mixed Weyl/boost anomaly, respectively fixed by the Kac-Moody and Virasoro central charges. Setting $dn=0$, as is relevant for the putative dual to gravity on WAdS$_3$, our analysis still applies, although now the Weyl/boost anomaly identically vanishes. The boost anomaly does not: under a boost $\psi$, the generating function varies by
\beq
\label{E:deltaWholo}
\delta_{\chi}W_k = \frac{k}{8\pi}\int d^2 x \sqrt{\gamma}\, \psi\,.
\eeq
For gravity on warped AdS$_3$ spacetimes, we identified the boost in the previous Subsection as a bulk local Lorentz rotation which falls off near the boundary as $v^0{}_1 = -\alpha \psi/r $. However, there does not seem to be any gravitational theory on WAdS$_3$ which realizes~\eqref{E:deltaWholo}. Any sensible bulk theory is locally Lorentz invariant up to a boundary term. So the anomaly~\eqref{E:deltaWholo} could only arise in one of two ways: from a bulk Chern-Simons term, or from a counterterm added in the course of holographically renormalizing the bulk. But the gravitational Chern-Simons does not lead to the variation~\eqref{E:deltaWholo}, and it is easy to check that there is no local boundary counterterm which realizes~\eqref{E:deltaWholo}. 

We conclude that the putative dual to gravity on WAdS$_3$ does not possess the warped anomalies of Section~\ref{S:anomalies}. This is somewhat puzzling, and we refer the reader to our comments on the matter in the Introduction. We suspect that this anomaly non-matching is an indirect consequence of the fact that the putative dual is also non-local, which we demonstrate presently.

\subsection{Semi-locality}

Consider a minimally coupled scalar $\varphi$ of mass $m^2$ on WAdS$_3$,
\beq
S_{matter} = - \int d^3x \sqrt{-g} \left(\frac{1}{2} (\partial\varphi)^2 + \frac{m^2}{2}\varphi^2\right)\,.
\eeq
In a pure WAdS$_3$ background we can solve for the asymptotic behavior of $\varphi$ near the boundary by Fourier transforming in the $x^{\pm}$ directions. Denoting
\beq
\varphi(x,r) = \int \frac{d\omega dk}{(2\pi)^2} e^{-i \omega x^- + i k x^+} \tilde{\varphi}(\omega,k,r)\,,
\eeq
a quick computation shows that $\tilde{\varphi}$ behaves at large $r$ as
\beq
\tilde{\varphi}(\omega,k,r) \sim a_1(\omega,k) r^{-1+\Delta(k)} + a_2(\omega,k) r^{-\Delta(k)}\,,
\eeq
with
\beq
\Delta(k) = \frac{1}{2}+\frac{\sqrt{1+4m^2L^2 + \frac{1-\alpha^2}{\alpha^2}k^2 }}{2}\,.
\eeq
By the usual holographic dictionary, the Fourier modes of $\varphi$ at fixed $k$ would be dual to an operator of dimension $\Delta(k)$. Two results immediately follow:
\begin{enumerate}
\item Since the dimension depends on the wavenumber of the mode, we see that $\varphi$ is not dual to a local operator in the putative dual.  Here is the non-locality we were after. This particular form of non-locality is the same as what one finds in fields propagating in the near-horizon AdS$_2\times\mathbb{R}^{d-1}$ geometry of an extremal Reissner-Nordstrom black brane, and so the dual to gravity on WAdS$_3$ is \emph{semi-local} in the same sense that the dual to gravity on AdS$_2\times \mathbb{R}^{d-1}$ is semi-local~\cite{Iqbal:2011in}.\footnote{The word ``semi'' is used because the dependence of the dimension $\Delta(k)$ on momentum at small $k$ is weak, $\Delta = \frac{1+\sqrt{1+4m^2L^2}}{2} + \mathcal{O}(k^2)$, and so the dual is approximately local over distances which are much longer than the WAdS scale $L$.} This should not be surprising, given that WAdS$_3$ is a line or circle fibration over AdS$_2$.
\item When the fiber is stretched $\alpha^2>1$, $\Delta(k)$ becomes complex for sufficiently large $|k|$. This indicates a dynamical instability whenever $\alpha^2>1$: matter fields on the background will condense, destroying the WAdS asymptotics. Further, modes with larger momentum condense more strongly than those with smaller momentum. Recall that there are spacelike WAdS$_3$ black holes only when $\alpha^2>1$. This result tells us that these black holes are always unstable.
\end{enumerate}

\subsection{Backreaction and NAdS$_2$ holography}

To conclude, a few words are in order about the AdS$_2$ factor in spacelike WAdS$_3$. Famously, AdS$_2$ cannot support finite energy excitations, and thus strictly speaking, AdS$_2$/CFT$_1$ holography does not exist. The backreaction of matter fields destroys the AdS$_2$ asymptotics. Instead one can formulate a NAdS$_2$/NCFT$_1$ correspondence~\cite{Almheiri:2014cka,Jensen:2016pah,Maldacena:2016upp,Engelsoy:2016xyb}, where the ``N'' stands for ``nearly,'' and the bulk geometry is (asymptotically) AdS$_2$ with a linear dilaton. 

There is a connection to spacelike WAdS$_3$. Suppose we compactify $x^+$ and take $\alpha^2<1$ so that matter fields are stable. Then another problem rears its head. Reducing to two dimensions, we obtain an AdS$_2$ spacetime with a constant electric field and a constant dilaton. Any matter fields will backreact strongly and destroy these asymptotics, so that spacelike WAdS$_3$ throats do not admit a decoupling limit, just like any AdS$_2$ throat with a finite volume transverse space. Consequently, these putative holographic WCFTs should be understood as nearly conformal theories, much the same way as the duals to NAdS$_2$ gravity. Further, they should be governed by the same low-energy ``Schwarzian action''~\cite{Maldacena:2016hyu} as NAdS$_2$ gravity~\cite{Jensen:2016pah,Maldacena:2016upp,Engelsoy:2016xyb}, supplemented with an additional phase field to account for the conserved momentum along the circle~\cite{Davison:2016ngz}.

\section*{Acknowledgments}

It is a pleasure to thank A.~Castro, T.~Hartman, D.~Hofman, J.~Hartong, N.~Iqbal, S.~Ross, B.~van Rees, and M.~Taylor for useful discussions. This work was supported in part by the US Department of Energy under Grant No. DE-SC0013682.

\begin{appendix}

\section{$\mathcal{W}$ and its variations}
\label{A:WA}

In Subsection~\ref{S:fromAnomToCentral} we computed the transformations of the WCFT currents $\mathcal{N}^{\mu}$ and $\mathcal{H}^{\mu}$ under a general coordinate transformation, Weyl rescaling, and boost. Therein we used the functional $\mathcal{W}$ which encodes the difference between $W$ evaluated on a background $(n_{\mu},h_{\nu})$ using coordinates $x^{\rho}$, and the generating functional on another background $(\hat{n}_{\mu},\hat{h}_{\nu})$ using coordinates $X^{\rho}$ related by an anomalous symmetry transformation,
\begin{equation*}
\hat{W}[\hat{n}_{\mu},\hat{h}_{\nu}';X^{\rho}] - W[n_{\mu},h_{\nu};x^{\rho}] = \mathcal{W}\,.
\end{equation*}
The background $(\hat{n}_{\mu},\hat{h}_{\nu})$ is related to $(n_{\mu},h_{\nu})$ by a coordinate transformation $x^{\mu}=x^{\mu}(X^{\nu})$ (with $g^{\mu}{}_{\nu} = \frac{\partial x^{\mu}}{\partial X^{\nu}}$), followed by a Weyl rescaling $\tau$, then a boost $\Psi$,
\begin{equation*}
\hat{n}_{\mu}(X) = e^{\tau}n_{\nu}g^{\nu}{}_{\mu}\,, \qquad \hat{h}_{\mu}(X) = (h_{\nu} - \Psi e^{\tau}n_{\nu})g^{\nu}{}_{\mu}\,.
\end{equation*}

In this Appendix we derive the expression~\eqref{E:WA} for $\mathcal{W}$ and compute its variations with respect to the background $(n_{\mu},h_{\nu})$.

The functional $\W$ is uniquely fixed by two constraints:
\begin{enumerate}
\item It vanishes when $x^{\mu}=X^{\mu}$, $\tau=0$, and $\Psi=0$.
\item It is a Wess-Zumino term for the anomalies. Under the transformation $\hat{\delta}_{\chi}$ which fixes $(n_{\mu},h_{\nu};x^{\rho})$ and implements an infinitesimal transformation on $(\hat{n}_{\mu},\hat{h}_{\nu};X^{\rho})$, that is
\begin{subequations}
\label{E:hatDelta}
\beq
\label{E:hatDelta1}
\hat{\delta}_{\chi} n_{\mu}(x) =\hat{ \delta}_{\chi}h_{\mu} (x)= 0\,,  \qquad \hat{\delta}_{\chi}x^{\mu} = 0\,,
\eeq
along with
\begin{align}
\begin{split}
\label{E:hatDelta2}
 \hat{\delta}_{\chi}X^{\mu} &= - \xi^{\mu}\,, 
\\
\hat{\delta}_{\chi}\hat{n}_{\mu}& = \pounds_{\xi}\hat{n}_{\mu} + \sigma \hat{n}_{\mu}\,, 
\\
\hat{\delta}_{\chi}\hat{h}_{\mu} &= \pounds_{\xi}\hat{h}_{\mu} - \psi \hat{n}_{\mu}\,,
\end{split}
\end{align}
\end{subequations}
$\mathcal{W}$ varies in such a way as to reproduce the anomalies of $\hat{W}$,
\beq
\label{E:reproduceAnomaly}
\hat{\delta}_{\chi}\W = \hat{\delta}_{\chi}\hat{W} \,,
\eeq
which given~\eqref{E:anomalies} we record here as
\begin{align}
\begin{split}
\label{E:hatAnomaly}
\hat{\delta}_{\chi}\hat{W}_{\tilde{k}} & = -\frac{\tilde{k}}{8\pi}\int d^2X \sqrt{\hat{\gamma}} \, \psi\,,
\\
\hat{\delta}_{\chi}\hat{W}_1 & = -\frac{\tilde{c}_1}{24\pi}\int d^2X\sqrt{\hat{\gamma}} \left( - \hat{N} \hat{v}^{\mu}\partial_{\mu} \sigma + \frac{\psi}{2}\hat{N}^2\right)\,,
\\
\hat{\delta}_{\chi}\hat{W}_2 & =- \frac{\tilde{c}_2}{192\pi}\int \left( \partial_{\mu}\xi^{\nu} d\hat{\Gamma}^{\mu}{}_{\nu} - d\psi \wedge \hat{\Gamma}^{\mu}{}_{\nu}\hat{w}^{\nu}\hat{n}_{\mu} + d\sigma \wedge \hat{\Gamma}^{\mu}{}_{\nu}\hat{v}^{\nu}\hat{n}_{\mu}\right)\,,
\end{split}
\end{align}
with
\beq
\hat{N} = N - \tau'\,, \qquad \hat{\Gamma}^{\mu}{}_{\nu} = \hat{v}^{\mu}d\hat{n}_{\nu} + \hat{w}^{\mu}d\hat{h}_{\nu}\,.
\eeq
\end{enumerate}

The hatted transformation~\eqref{E:hatDelta} is equivalent to~\eqref{E:hatDelta1} and 
\beq
\hat{\delta}_{\chi} g^{\mu}{}_{\nu} = g^{\mu}{}_{\rho}\partial_{\nu}\xi^{\rho}\,, \qquad \hat{\delta}_{\chi} \tau = \sigma\,, \qquad \hat{\delta}_{\chi} \Psi = \psi - \sigma \Psi\,.
\eeq

We now show that the expression~\eqref{E:WA} for $\W$ does the job. To proceed, we split $\W$ into three parts, one for each anomaly,
\beq
\W = \W_{\tilde{k}} + \W_1 + \W_2\,,
\eeq
with 
\beq
\W_{\tilde{k}} = -\frac{\tilde{k}}{8\pi}\int d^2x \sqrt{\gamma} \,\Psi e^{\tau} =- \frac{\tilde{k}}{8\pi}\int d^2X \sqrt{\hat{\gamma}}\,\Psi\,, 
\eeq
accounting for the boost anomaly~\eqref{E:boostAnomaly}, and
\beq
\W_1 = -\frac{\tilde{c}_1}{24\pi}\int d^2x \sqrt{\gamma}\left\{ - N \dot{\tau} + \frac{1}{2}\dot{\tau}\tau' + \frac{\Psi e^{\tau}}{2}(N-\tau')^2\right\}
\eeq
the mixed boost/Weyl anomaly~\eqref{E:mixedAnomaly}. The term that accounts for the gravitational anomaly~\eqref{E:gravAnomaly} contains a non-abelian WZ term and so cannot be written covariantly in two dimensions. As we described in Subsection~\ref{S:fromAnomToCentral}, that WZ term includes an integral on a three-manifold $\mathcal{N}$ with the spacetime manifold $\mathcal{M}$ as its boundary. We find
\begin{align}
\nonumber
\W_2 & =  \frac{\tilde{c}_2}{192\pi}\left\{ \frac{1}{3} \int_{\mathcal{N}}\text{tr}\left(dg \, g^{-1}\right)^3 + \int dg^{\mu}{}_{\nu} (g^{-1})^{\nu}{}_{\rho} \wedge \Gamma^{\rho}{}_{\mu} \right\}
\\
\label{E:finalW2}
& \qquad- \frac{\tilde{c}_2}{192\pi}\int \left\{ \left(v^{\nu}+\Psi e^{\tau}w^{\nu}\right)n_{\mu}d\tau - w^{\nu}n_{\mu}d\left( \Psi e^{\tau}\right)\right\} \wedge \left( \Gamma^{\mu}{}_{\nu}  + dg^{\mu}{}_{\rho}(g^{-1})^{\rho}{}_{\nu}\right)
\\
\nonumber
& = \frac{\tilde{c}_2}{192\pi}\Big\{ \frac{1}{3}\int_{\mathcal{N}}\text{tr}\left(dg\, g^{-1}\right)^3 - \int \left( g^{\mu}{}_{\rho}\hat{\Gamma}^{\rho}{}_{\sigma}(g^{-1})^{\sigma}{}_{\nu} - dg^{\mu}{}_{\rho}(g^{-1})^{\rho}{}_{\nu}\right)\wedge \left( \Gamma^{\nu}{}_{\mu} + dg^{\nu}{}_{\alpha}(g^{-1})^{\alpha}{}_{\mu}\right)\Big\}\,.
\end{align}

Each of these satisfies the first requirement: $\mathcal{W}$ vanishes when $g^{\mu}{}_{\nu}=\delta^{\mu}{}_{\nu}\,, \tau=\Psi=0$. So it remains to check that these functionals satisfy~\eqref{E:reproduceAnomaly} with~\eqref{E:hatAnomaly}. We begin with the pure boost anomaly. Its variation is
\beq
\hat{\delta}_{\chi}\W_{\tilde{k}} = -\frac{\tilde{k}}{8\pi}\int d^2X \sqrt{\hat{\gamma}} \, \psi \,,
\eeq
as it ought. The variation of $\mathcal{W}_1$ gives (using $\hat{N} = N-\tau'$)
\begin{align}
\begin{split}
\hat{\delta}_{\chi} \W_1 &= -\frac{\tilde{c}_1}{24\pi}\int d^2x \sqrt{\gamma} \left\{ - (N-\tau')(v^{\mu} + \Psi e^{\tau}w^{\mu})\partial_{\mu}\sigma  + \frac{\psi e^{\tau}}{2}(N-\tau')^2\right\}
\\
& = -\frac{\tilde{c}_1}{24\pi}\int d^2X\sqrt{\hat{\gamma}} \left\{ - \hat{N} \hat{v}^{\mu}\partial_{\mu}\sigma + \frac{\psi}{2}\hat{N}^2\right\}\,.
\end{split}
\end{align}
Finally, using
\beq
\hat{\delta}{}_{\chi} \left( dg^{\mu}{}_{\rho} (g^{-1})^{\rho}{}_{\nu}\right) = g^{\mu}{}_{\rho}\, d\partial_{\sigma}\xi^{\rho}\, (g^{-1})^{\sigma}{}_{\nu}\,,
\eeq
and
\beq
\hat{\Gamma}^{\mu}{}_{\nu} = (g^{-1})^{\mu}{}_{\rho}\Gamma^{\rho}{}_{\sigma}g^{\sigma}{}_{\nu} + (g^{-1})^{\mu}{}_{\rho}dg^{\rho}{}_{\nu} + (g^{-1})^{\mu}{}_{\rho}\Big( (v^{\rho}+\Psi e^{\tau}w^{\rho})n_{\sigma}d\tau - w^{\rho}n_{\sigma} d\left( \Psi e^{\tau}\right)\Big)g^{\sigma}{}_{\nu}\,,
\eeq
we obtain the variation of $\W_2$ (up to a now-ubiquitous boundary term),
\begin{align}
\begin{split}
\hat{\delta}_{\chi} \W_2 & = \frac{\tilde{c}_2}{192\pi}\Big\{ \int d\partial_{\mu}\xi^{\nu} \wedge \left( (g^{-1})^{\mu}{}_{\rho}\Gamma^{\rho}{}_{\sigma} g^{\sigma}{}_{\nu} + (g^{-1})^{\mu}{}_{\rho}dg^{\rho}{}_{\nu}\right) 
\\
& \qquad  \qquad \quad + d\partial_{\mu}\xi^{\nu} \wedge \Big( (v^{\mu} + \Psi e^{\tau} w^{\mu})n_{\nu}d\tau - w^{\mu}n_{\nu} d\left( \Psi e^{\tau}\right)\Big)
\\
& \qquad \qquad \qquad \quad + d\psi \wedge e^{\tau} n_{\mu}\left( \Gamma^{\mu}{}_{\nu} + dg^{\mu}{}_{\rho}(g^{-1})^{\rho}{}_{\nu}\right) w^{\nu}
\\
& \qquad \qquad\qquad \qquad \quad - d\sigma \wedge n_{\mu}\left( \Gamma^{\mu}{}_{\nu} +dg^{\mu}{}_{\rho}(g^{-1})^{\rho}{}_{\nu}\right) \left( v^{\nu} + \Psi e^{\tau}w^{\nu}\right)\Big\}
\\
& = - \frac{\tilde{c}_2}{192\pi}\int \left( \partial_{\mu}\xi^{\nu}d\hat{\Gamma}^{\mu}{}_{\nu} - d\psi \wedge \hat{\Gamma}^{\mu}{}_{\nu}\hat{w}^{\nu}\hat{n}_{\mu} + d\sigma \hat{\Gamma}^{\mu}{}_{\nu}\hat{v}^{\nu}\hat{n}_{\mu}\right)\,.
\end{split}
\end{align}

Having computed the functional $\W$, we compute its variations with respect to the background fields $(n_{\mu},h_{\nu})$ in order to obtain the ``anomalous currents'' $(\mathcal{N}^{\mu}_A,\mathcal{H}^{\nu}_A)$. We also decompose these into three parts, one for each anomaly:
\beq
\mathcal{N}_A^{\mu} = \mathcal{N}_{\tilde{k}}^{\mu} + \mathcal{N}_1^{\mu} + \mathcal{N}_2^{\mu} \,, \qquad \mathcal{H}_A^{\mu} = \mathcal{H}_{\tilde{k}}^{\mu} + \mathcal{H}_1^{\mu} + \mathcal{H}_2^{\mu}\,.
\eeq
The variations of $\W_{\tilde{k}}$ give
\begin{align}
\begin{split}
\mathcal{N}_{\tilde{k}}^{\mu} &= -\frac{\tilde{k}}{4} \Psi e^{\tau} v^{\mu}\,,
\\
 \mathcal{H}_{\tilde{k}}^{\mu}& = -\frac{\tilde{k}}{4} \Psi e^{\tau} w^{\mu}\,,
\end{split}
\end{align}
and the variations of $\W_1$ give
\begin{align}
\begin{split}
\mathcal{N}_1^{\mu} & =-\frac{ \tilde{c}_1}{12}\left\{  \dot{\tau}Nv^{\mu} - \frac{\dot{\tau}^2}{2}w^{\mu} + \Psi e^{\tau} \left(  \frac{\tau'^2-N^2}{2}v^{\mu} + \dot{\tau}(N-\tau')w^{\mu} \right)\right\}
\\
& \qquad \qquad -\frac{\tilde{c}_1}{12} \varepsilon^{\mu\nu}\partial_{\nu}\left( \Psi e^{\tau}(N-\tau')-\dot{\tau}\right)\,,
\\
\mathcal{H}_1^{\mu} & =-\frac{ \tilde{c}_1}{12} \left\{ \tau' N v^{\mu} - \frac{\tau'^2}{2}v^{\mu} -\frac{(N-\tau')^2}{2}\Psi w^{\mu}\right\}\,.
\end{split}
\end{align}

The variations of $\W_2$ are rather complex. Rather than present them, let us argue that they vanish in the flat ``Cartesian'' background $n = dx^-, \,h=dx^+$ for the warped conformal transformations~\eqref{E:WCFTsymmetries2}. We begin with the expression for $\W_2$ given in the last line of~\eqref{E:finalW2}. In this background and choice of coordinates, $dn_{\mu}=dh_{\nu}=0$ and $\Gamma^{\mu}{}_{\nu}=0$. Similarly, in the transformed background we have $d\hat{n}_{\mu}=d\hat{n}_{\nu}=0$ and $\hat{\Gamma}^{\mu}{}_{\nu}=0$. Moreover, the variations of $\Gamma^{\mu}{}_{\nu}$ and $\hat{\Gamma}^{\mu}{}_{\nu}$ are total derivatives:
\beq
\delta_{\rm flat}\Gamma^{\mu}{}_{\nu} =d\big(v^{\mu}\delta n_{\nu}  + w^{\mu}\delta h_{\nu}\big)\,, \qquad \delta_{\rm flat} \hat{\Gamma}^{\mu}{}_{\nu}  = d \big( \hat{v}^{\mu}\delta \hat{n}_{\nu} + \hat{w}^{\mu}\delta \hat{h}_{\nu}\big)\,,
\eeq
where the ``flat'' subscript that these equalities hold in the flat background with ``Cartesian'' coordinates. After an integration by parts, the variation of the expression for $\W_2$ in the last line of~\eqref{E:finalW2} is proportional to
\beq
dg^{\mu}{}_{\nu} \wedge d(g^{-1})^{\nu}{}_{\rho}\,.
\eeq
However, given that $dg^{\mu}{}_{\nu} \propto dX^-$ for the transformation at hand, we see that this wedge product vanishes and so the functional variation of $\W_2$ does too. So, for the warped conformal symmetries~\eqref{E:WCFTsymmetries2} and flat background, we find
\beq
\mathcal{N}_2^{\mu} =_{\rm flat} 0\,, \qquad \mathcal{H}_2^{\mu}=_{\rm flat} 0\,.
\eeq

\end{appendix}

\bibliography{wRefs}

\providecommand{\href}[2]{#2}\begingroup\raggedright\begin{thebibliography}{10}

\bibitem{Guica:2008mu}
M.~Guica, T.~Hartman, W.~Song, and A.~Strominger, {\it {The Kerr/CFT
  Correspondence}},  {\em Phys. Rev.} {\bf D80} (2009) 124008,
  [\href{http://xxx.lanl.gov/abs/0809.4266}{{\tt arXiv:0809.4266}}].

\bibitem{Compere:2008cv}
G.~Compere and S.~Detournay, {\it {Semi-classical central charge in
  topologically massive gravity}},  {\em Class. Quant. Grav.} {\bf 26} (2009)
  012001, [\href{http://xxx.lanl.gov/abs/0808.1911}{{\tt arXiv:0808.1911}}].
  [Erratum: Class. Quant. Grav.26,139801(2009)].

\bibitem{Anninos:2008fx}
D.~Anninos, W.~Li, M.~Padi, W.~Song, and A.~Strominger, {\it {Warped AdS(3)
  Black Holes}},  {\em JHEP} {\bf 03} (2009) 130,
  [\href{http://xxx.lanl.gov/abs/0807.3040}{{\tt arXiv:0807.3040}}].

\bibitem{Anninos:2008qb}
D.~Anninos, {\it {Hopfing and Puffing Warped Anti-de Sitter Space}},  {\em
  JHEP} {\bf 09} (2009) 075, [\href{http://xxx.lanl.gov/abs/0809.2433}{{\tt
  arXiv:0809.2433}}].

\bibitem{Guica:2010sw}
M.~Guica, K.~Skenderis, M.~Taylor, and B.~C. van Rees, {\it {Holography for
  Schrodinger backgrounds}},  {\em JHEP} {\bf 02} (2011) 056,
  [\href{http://xxx.lanl.gov/abs/1008.1991}{{\tt arXiv:1008.1991}}].

\bibitem{Hofman:2011zj}
D.~M. Hofman and A.~Strominger, {\it {Chiral Scale and Conformal Invariance in
  2D Quantum Field Theory}},  {\em Phys. Rev. Lett.} {\bf 107} (2011) 161601,
  [\href{http://xxx.lanl.gov/abs/1107.2917}{{\tt arXiv:1107.2917}}].

\bibitem{ElShowk:2011cm}
S.~El-Showk and M.~Guica, {\it {Kerr/CFT, dipole theories and nonrelativistic
  CFTs}},  {\em JHEP} {\bf 12} (2012) 009,
  [\href{http://xxx.lanl.gov/abs/1108.6091}{{\tt arXiv:1108.6091}}].

\bibitem{Song:2011sr}
W.~Song and A.~Strominger, {\it {Warped AdS3/Dipole-CFT Duality}},  {\em JHEP}
  {\bf 05} (2012) 120, [\href{http://xxx.lanl.gov/abs/1109.0544}{{\tt
  arXiv:1109.0544}}].

\bibitem{Guica:2011ia}
M.~Guica, {\it {A Fefferman-Graham-Like Expansion for Null Warped AdS(3)}},
  {\em JHEP} {\bf 12} (2012) 084,
  [\href{http://xxx.lanl.gov/abs/1111.6978}{{\tt arXiv:1111.6978}}].

\bibitem{Azeyanagi:2012zd}
T.~Azeyanagi, D.~M. Hofman, W.~Song, and A.~Strominger, {\it {The Spectrum of
  Strings on Warped AdS$_3 x S^3$}},  {\em JHEP} {\bf 04} (2013) 078,
  [\href{http://xxx.lanl.gov/abs/1207.5050}{{\tt arXiv:1207.5050}}].

\bibitem{Detournay:2012pc}
S.~Detournay, T.~Hartman, and D.~M. Hofman, {\it {Warped Conformal Field
  Theory}},  {\em Phys. Rev.} {\bf D86} (2012) 124018,
  [\href{http://xxx.lanl.gov/abs/1210.0539}{{\tt arXiv:1210.0539}}].

\bibitem{Compere:2013bya}
G.~Compere, W.~Song, and A.~Strominger, {\it {New Boundary Conditions for
  AdS3}},  {\em JHEP} {\bf 05} (2013) 152,
  [\href{http://xxx.lanl.gov/abs/1303.2662}{{\tt arXiv:1303.2662}}].

\bibitem{Anninos:2013nja}
D.~Anninos, J.~Samani, and E.~Shaghoulian, {\it {Warped Entanglement Entropy}},
   {\em JHEP} {\bf 02} (2014) 118,
  [\href{http://xxx.lanl.gov/abs/1309.2579}{{\tt arXiv:1309.2579}}].

\bibitem{Hofman:2014loa}
D.~M. Hofman and B.~Rollier, {\it {Warped Conformal Field Theory as Lower Spin
  Gravity}},  {\em Nucl. Phys.} {\bf B897} (2015) 1--38,
  [\href{http://xxx.lanl.gov/abs/1411.0672}{{\tt arXiv:1411.0672}}].

\bibitem{Hartong:2015xda}
J.~Hartong, {\it {Gauging the Carroll Algebra and Ultra-Relativistic Gravity}},
   {\em JHEP} {\bf 08} (2015) 069,
  [\href{http://xxx.lanl.gov/abs/1505.0501}{{\tt arXiv:1505.0501}}].

\bibitem{Castro:2015uaa}
A.~Castro, D.~M. Hofman, and G.~Sarosi, {\it {Warped Weyl fermion partition
  functions}},  {\em JHEP} {\bf 11} (2015) 129,
  [\href{http://xxx.lanl.gov/abs/1508.0630}{{\tt arXiv:1508.0630}}].

\bibitem{Castro:2015csg}
A.~Castro, D.~M. Hofman, and N.~Iqbal, {\it {Entanglement Entropy in Warped
  Conformal Field Theories}},  {\em JHEP} {\bf 02} (2016) 033,
  [\href{http://xxx.lanl.gov/abs/1511.0070}{{\tt arXiv:1511.0070}}].

\bibitem{Castro:2017mfj}
A.~Castro, C.~Keeler, and P.~Szepietowski, {\it {Tweaking one-loop determinants
  in AdS$_{3}$}},  {\em JHEP} {\bf 10} (2017) 070,
  [\href{http://xxx.lanl.gov/abs/1707.0624}{{\tt arXiv:1707.0624}}].

\bibitem{Guica:2017lia}
M.~Guica, {\it {An integrable Lorentz-breaking deformation of two-dimensional
  CFTs}},  \href{http://xxx.lanl.gov/abs/1710.0841}{{\tt arXiv:1710.0841}}.

\bibitem{Geracie:2014nka}
M.~Geracie, D.~T. Son, C.~Wu, and S.-F. Wu, {\it {Spacetime Symmetries of the
  Quantum Hall Effect}},  {\em Phys. Rev.} {\bf D91} (2015) 045030,
  [\href{http://xxx.lanl.gov/abs/1407.1252}{{\tt arXiv:1407.1252}}].

\bibitem{Christensen:2013rfa}
M.~H. Christensen, J.~Hartong, N.~A. Obers, and B.~Rollier, {\it {Boundary
  Stress-Energy Tensor and Newton-Cartan Geometry in Lifshitz Holography}},
  {\em JHEP} {\bf 01} (2014) 057,
  [\href{http://xxx.lanl.gov/abs/1311.6471}{{\tt arXiv:1311.6471}}].

\bibitem{Jensen:2012jh}
K.~Jensen, M.~Kaminski, P.~Kovtun, R.~Meyer, A.~Ritz, and A.~Yarom, {\it
  {Towards hydrodynamics without an entropy current}},  {\em Phys. Rev. Lett.}
  {\bf 109} (2012) 101601, [\href{http://xxx.lanl.gov/abs/1203.3556}{{\tt
  arXiv:1203.3556}}].

\bibitem{Banerjee:2012iz}
N.~Banerjee, J.~Bhattacharya, S.~Bhattacharyya, S.~Jain, S.~Minwalla, and
  T.~Sharma, {\it {Constraints on Fluid Dynamics from Equilibrium Partition
  Functions}},  {\em JHEP} {\bf 09} (2012) 046,
  [\href{http://xxx.lanl.gov/abs/1203.3544}{{\tt arXiv:1203.3544}}].

\bibitem{Iqbal:2011in}
N.~Iqbal, H.~Liu, and M.~Mezei, {\it {Semi-local quantum liquids}},  {\em JHEP}
  {\bf 04} (2012) 086, [\href{http://xxx.lanl.gov/abs/1105.4621}{{\tt
  arXiv:1105.4621}}].

\bibitem{Jensen:2014aia}
K.~Jensen, {\it {On the coupling of Galilean-invariant field theories to curved
  spacetime}},  \href{http://xxx.lanl.gov/abs/1408.6855}{{\tt
  arXiv:1408.6855}}.

\bibitem{Geracie:2015dea}
M.~Geracie, K.~Prabhu, and M.~M. Roberts, {\it {Curved non-relativistic
  spacetimes, Newtonian gravitation and massive matter}},  {\em J. Math. Phys.}
  {\bf 56} (2015), no.~10 103505,
  [\href{http://xxx.lanl.gov/abs/1503.0268}{{\tt arXiv:1503.0268}}].

\bibitem{Hartong:2015wxa}
J.~Hartong, E.~Kiritsis, and N.~A. Obers, {\it {Field Theory on Newton-Cartan
  Backgrounds and Symmetries of the Lifshitz Vacuum}},  {\em JHEP} {\bf 08}
  (2015) 006, [\href{http://xxx.lanl.gov/abs/1502.0022}{{\tt
  arXiv:1502.0022}}].

\bibitem{Hartong:2015usd}
J.~Hartong, {\it {Holographic Reconstruction of 3D Flat Space-Time}},
  \href{http://xxx.lanl.gov/abs/1511.0138}{{\tt arXiv:1511.0138}}.

\bibitem{Jensen:2014hqa}
K.~Jensen, {\it {Anomalies for Galilean fields}},
  \href{http://xxx.lanl.gov/abs/1412.7750}{{\tt arXiv:1412.7750}}.

\bibitem{Arav:2016xjc}
I.~Arav, S.~Chapman, and Y.~Oz, {\it {Non-Relativistic Scale Anomalies}},  {\em
  JHEP} {\bf 06} (2016) 158, [\href{http://xxx.lanl.gov/abs/1601.0679}{{\tt
  arXiv:1601.0679}}].

\bibitem{Pal:2017ntk}
S.~Pal and B.~Grinstein, {\it {On the Heat Kernel and Weyl Anomaly of
  Schr\"odinger invariant theory}},
  \href{http://xxx.lanl.gov/abs/1703.0298}{{\tt arXiv:1703.0298}}.

\bibitem{Auzzi:2017jry}
R.~Auzzi, S.~Baiguera, and G.~Nardelli, {\it {Trace anomaly for
  non-relativistic fermions}},  {\em JHEP} {\bf 08} (2017) 042,
  [\href{http://xxx.lanl.gov/abs/1705.0222}{{\tt arXiv:1705.0222}}].

\bibitem{Fernandes:2017nvx}
K.~Fernandes and A.~Mitra, {\it {Gravitational anomalies on the Newton-Cartan
  background}},  {\em Phys. Rev.} {\bf D96} (2017), no.~8 085003,
  [\href{http://xxx.lanl.gov/abs/1703.0916}{{\tt arXiv:1703.0916}}].

\bibitem{Gromov:2015fda}
A.~Gromov, K.~Jensen, and A.~G. Abanov, {\it {Boundary effective action for
  quantum Hall states}},  {\em Phys. Rev. Lett.} {\bf 116} (2016), no.~12
  126802, [\href{http://xxx.lanl.gov/abs/1506.0717}{{\tt arXiv:1506.0717}}].

\bibitem{Bardeen:1984pm}
W.~A. Bardeen and B.~Zumino, {\it {Consistent and Covariant Anomalies in Gauge
  and Gravitational Theories}},  {\em Nucl. Phys.} {\bf B244} (1984) 421.

\bibitem{Jensen:2014ama}
K.~Jensen, {\it {Aspects of hot Galilean field theory}},  {\em JHEP} {\bf 04}
  (2015) 123, [\href{http://xxx.lanl.gov/abs/1411.7024}{{\tt
  arXiv:1411.7024}}].

\bibitem{Jensen:2014wha}
K.~Jensen and A.~Karch, {\it {Revisiting non-relativistic limits}},  {\em JHEP}
  {\bf 04} (2015) 155, [\href{http://xxx.lanl.gov/abs/1412.2738}{{\tt
  arXiv:1412.2738}}].

\bibitem{Jensen:2013kka}
K.~Jensen, R.~Loganayagam, and A.~Yarom, {\it {Anomaly inflow and thermal
  equilibrium}},  {\em JHEP} {\bf 05} (2014) 134,
  [\href{http://xxx.lanl.gov/abs/1310.7024}{{\tt arXiv:1310.7024}}].

\bibitem{Jensen:2013rga}
K.~Jensen, R.~Loganayagam, and A.~Yarom, {\it {Chern-Simons terms from thermal
  circles and anomalies}},  {\em JHEP} {\bf 05} (2014) 110,
  [\href{http://xxx.lanl.gov/abs/1311.2935}{{\tt arXiv:1311.2935}}].

\bibitem{Jensen:2012kj}
K.~Jensen, R.~Loganayagam, and A.~Yarom, {\it {Thermodynamics, gravitational
  anomalies and cones}},  {\em JHEP} {\bf 02} (2013) 088,
  [\href{http://xxx.lanl.gov/abs/1207.5824}{{\tt arXiv:1207.5824}}].

\bibitem{Golkar:2015oxw}
S.~Golkar and S.~Sethi, {\it {Global Anomalies and Effective Field Theory}},
  \href{http://xxx.lanl.gov/abs/1512.0260}{{\tt arXiv:1512.0260}}.

\bibitem{Chowdhury:2016cmh}
S.~D. Chowdhury and J.~R. David, {\it {Global gravitational anomalies and
  transport}},  \href{http://xxx.lanl.gov/abs/1604.0500}{{\tt
  arXiv:1604.0500}}.

\bibitem{Glorioso:2017lcn}
P.~Glorioso, H.~Liu, and S.~Rajagopal, {\it {Global Anomalies, Discrete
  Symmetries, and Hydrodynamic Effective Actions}},
  \href{http://xxx.lanl.gov/abs/1710.0376}{{\tt arXiv:1710.0376}}.

\bibitem{Moussa:2003fc}
K.~A. Moussa, G.~Clement, and C.~Leygnac, {\it {The Black holes of
  topologically massive gravity}},  {\em Class. Quant. Grav.} {\bf 20} (2003)
  L277--L283, [\href{http://xxx.lanl.gov/abs/gr-qc/0303042}{{\tt
  gr-qc/0303042}}].

\bibitem{Bardeen:1999px}
J.~M. Bardeen and G.~T. Horowitz, {\it {The Extreme Kerr throat geometry: A
  Vacuum analog of AdS(2) x S**2}},  {\em Phys. Rev.} {\bf D60} (1999) 104030,
  [\href{http://xxx.lanl.gov/abs/hep-th/9905099}{{\tt hep-th/9905099}}].

\bibitem{Bengtsson:2005zj}
I.~Bengtsson and P.~Sandin, {\it {Anti de Sitter space, squashed and
  stretched}},  {\em Class. Quant. Grav.} {\bf 23} (2006) 971--986,
  [\href{http://xxx.lanl.gov/abs/gr-qc/0509076}{{\tt gr-qc/0509076}}].

\bibitem{Imbimbo:1999bj}
C.~Imbimbo, A.~Schwimmer, S.~Theisen, and S.~Yankielowicz, {\it
  {Diffeomorphisms and holographic anomalies}},  {\em Class. Quant. Grav.} {\bf
  17} (2000) 1129--1138, [\href{http://xxx.lanl.gov/abs/hep-th/9910267}{{\tt
  hep-th/9910267}}].

\bibitem{Almheiri:2014cka}
A.~Almheiri and J.~Polchinski, {\it {Models of AdS$_{2}$ backreaction and
  holography}},  {\em JHEP} {\bf 11} (2015) 014,
  [\href{http://xxx.lanl.gov/abs/1402.6334}{{\tt arXiv:1402.6334}}].

\bibitem{Jensen:2016pah}
K.~Jensen, {\it {Chaos in AdS$_2$ Holography}},  {\em Phys. Rev. Lett.} {\bf
  117} (2016), no.~11 111601, [\href{http://xxx.lanl.gov/abs/1605.0609}{{\tt
  arXiv:1605.0609}}].

\bibitem{Maldacena:2016upp}
J.~Maldacena, D.~Stanford, and Z.~Yang, {\it {Conformal symmetry and its
  breaking in two dimensional Nearly Anti-de-Sitter space}},  {\em PTEP} {\bf
  2016} (2016), no.~12 12C104, [\href{http://xxx.lanl.gov/abs/1606.0185}{{\tt
  arXiv:1606.0185}}].

\bibitem{Engelsoy:2016xyb}
J.~Engelsoy, T.~G. Mertens, and H.~Verlinde, {\it {An investigation of
  AdS$_{2}$ backreaction and holography}},  {\em JHEP} {\bf 07} (2016) 139,
  [\href{http://xxx.lanl.gov/abs/1606.0343}{{\tt arXiv:1606.0343}}].

\bibitem{Maldacena:2016hyu}
J.~Maldacena and D.~Stanford, {\it {Remarks on the Sachdev-Ye-Kitaev model}},
  {\em Phys. Rev.} {\bf D94} (2016), no.~10 106002,
  [\href{http://xxx.lanl.gov/abs/1604.0781}{{\tt arXiv:1604.0781}}].

\bibitem{Davison:2016ngz}
R.~A. Davison, W.~Fu, A.~Georges, Y.~Gu, K.~Jensen, and S.~Sachdev, {\it
  {Thermoelectric transport in disordered metals without quasiparticles: The
  Sachdev-Ye-Kitaev models and holography}},  {\em Phys. Rev.} {\bf B95}
  (2017), no.~15 155131, [\href{http://xxx.lanl.gov/abs/1612.0084}{{\tt
  arXiv:1612.0084}}].

\end{thebibliography}\endgroup
\bibliographystyle{JHEP}

\end{document}